\documentclass[journal]{IEEEtran}
\usepackage{times}
\usepackage{epsfig}
\usepackage{graphicx}
\usepackage{amsmath}
\usepackage{amssymb}
\usepackage{cite}
\usepackage{hyperref}
\usepackage{orcidlink}

\begin{document}
\title{AutoAtlas: Neural Network for 3D Unsupervised Partitioning and Representation Learning}

\author{K.~Aditya~Mohan\textsuperscript{\orcidlink{0000-0002-0921-6559}},~\IEEEmembership{Senior Member,~IEEE,}~Alan~D.~Kaplan\textsuperscript{\orcidlink{0000-0002-4999-7687}},~\IEEEmembership{Member,~IEEE}
\thanks{K.~A.~Mohan is with the Computational Engineering Division (CED)
at Lawrence Livermore National Laboratory, Livermore, CA, 94551 USA.
E-mail: mohan3@llnl.gov, adityakadri@gmail.com.}
\thanks{A.~D.~Kaplan is with the Computational Engineering Division (CED)
at Lawrence Livermore National Laboratory, Livermore, CA, 94551 USA. E-mail: kaplan7@llnl.gov.}%
\thanks{We also include supporting files in the form of videos.}}%


\maketitle

\begin{abstract}
We present a novel neural network architecture called AutoAtlas for fully unsupervised partitioning and representation learning of 3D brain Magnetic Resonance Imaging (MRI) volumes.
AutoAtlas consists of two neural network components: one neural network to perform multi-label partitioning based on local texture in the volume, and a second neural network to compress the information contained within each partition.
We train both of these components simultaneously by optimizing a loss function that is designed to promote accurate reconstruction of each partition, while   encouraging spatially smooth and contiguous partitioning, and discouraging relatively small partitions.
We show that the partitions adapt to the subject specific structural variations of brain tissue
while consistently appearing at similar spatial locations across subjects. AutoAtlas also produces very low dimensional features that represent local texture of each partition.
We demonstrate prediction of metadata associated with each subject using the derived feature representations and compare the results to prediction using features derived from FreeSurfer anatomical parcellation.
Since our features are intrinsically linked to distinct partitions, we can then map values of interest, such as partition-specific feature importance scores onto the brain for visualization.
\end{abstract}

\begin{IEEEkeywords}
Brain Imaging, MRI, Representation Learning, Deep Learning, CNN.
\end{IEEEkeywords}

\section{Introduction}
\IEEEPARstart{R}epresentation learning is an important task in data analysis and machine learning.
The goal of representation learning is to find a compressed, lower dimensional version of the data that is suitable for a particular task, such as prediction or pattern discovery \cite{Bengio2013-ao}.
Common examples include Principal Component Analysis (PCA), nonlinear kernel representation, and autoencoders (as computed by Neural Networks), see e.g. \cite{Fukunaga2013-wk, Shawe-Taylor2004-gx, Goodfellow2016-xu}.
In addition, it is often desired to compute representations that contain meaningful semantic interpretation for the specific applications.
In natural images, this could include features, or dimensions, in the representation that correspond to elements such as pose, background, or lighting conditions (see e.g. \cite{Chen2016-kg, Tran2017-zs}).

\subsection{Representation Learning for Volumes}
A particularly difficult challenge is performing representation learning for 3-dimensional (3D) volumes, such as Magnetic Resonance Imaging (MRI) brain images.
First is the challenge of scale and resulting complexity of 3-dimensional data.
Treating 3D data as a series of 2D slices is often undesired, since a particular choice of spatial slice orientation is arbitrary.
Secondly, for medical imagery such as MRI, we are often concerned about spatially localizing the derived representation.
A representation that maintains the 3D organization and is spatially sensitive would enable a common vocabulary of the representation across individuals (e.g. features related to the same tissue type such as bone can then be compared).

The brain as a complex organ adds another difficult challenge because of its detailed anatomical organization.
With respect to tissue structure, the brain contains cortical gray matter, white matter, subcortical gray matter, and the ventricles.
We exclude the bone of the skull in this list because tools exist to extract the brain from the skull within MRI.
Beyond the tissue types, regions of the cortical and subcortical gray matter are known to be tuned to specific functional tasks, for example, related to performing motor or visual tasks \cite{Eickhoff2018-wc}.
Indeed, the interplay between the spatial organization of the brain and its function is a central theme in neuroscience, where it is known that the anatomical regions in the brain perform distinct tasks \cite{Bassett2017-cb}.
Because of this, invariance to spatial location of learned image features may be an undesired effect for representation learning of brain imagery.
All of this organization can be incorporated in representations computed from MRI.
One key challenge, therefore, is the construction of representations which are sensitive to the spatial layout of the brain.

In this paper, we present AutoAtlas, which is a data-driven and fully unsupervised approach for constructing representations of volumetric imagery that simultaneously performs partitioning of the volume and feature extraction of each partition.
This is accomplished by encouraging the representation to adhere to spatial constraints in the loss function.
The result is a novel approach to representation learning for 3D volumes that directly associates spatial partitions with learned representations, and we demonstrate the utility of this approach in an empirical prediction task. 
We utilize data from the Human Connectome Project to train the models, which has released a dataset of healthy young adults (HCP-YA) that includes approximately 1,200 subjects along with static measures such as prior health history, emotional processing, cognitive performance, and motor skills \cite{Van_Essen2013-xf, Glasser2016-kr, Smith2013-vw, Barch2013-wy, Van_Essen2012-cf}. 
We use the output of the minimal processing pipeline as specified by the HCP-YA documentation \cite{glasser2013minimal}.  
For this work, we utilized T1 MRI volumes in native volume space. 
An IRB exemption was obtained through the LLNL Institutional Review Board for use of the HCP-YA data.

AutoAtlas simultaneously learns partitions that correspond across subjects and features associated with each partition.
This is accomplished by a connected pair of neural network architectures: one that predicts partition labels in the input volume, and another that compresses the information within each partition.
The approach does not require any manual labeling or secondary partitioning of the volumes and therefore operates in a fully unsupervised setting.
We show examples of the resulting partitions and quantify its stability across patients by anchoring to tissue segmentations.

\subsection{Prediction Using MRI}
Data-driven analysis and machine learning has shown promise in finding associations between brain imagery and predictive targets \cite{Landhuis2017-mv, Lemm2011-an}.
These approaches rely on large collections of imagery and are trained to predict disease states or behavioral characteristics.
This type of prediction using brain imagery is an important problem in computational neuroscience with broad applications in clinical and scientific domains \cite{Landhuis2017-mv}.
The advantage of separating representation learning and prediction is of special importance to many applications that require multiple predictions because the generation of distinct representations for each separate predictive task obstructs spatially derived insights that may be present across predictive tasks.


The novel contributions of this paper are as follows -
\begin{itemize}
\item Neural network architecture called AutoAtlas for fully unsupervised data-driven partitioning and representation learning of 3D volumes without the use of any prior knowledge.
\item Loss function called Neighborhood Label Similarity (NLS) loss that encourages AutoAtlas to produce partitions where each voxel is forced to predict a single partition label with high confidence. The NLS loss is also responsible for producing smooth and contiguous partitions.
\item Loss function called Anti-Devouring (AD) loss which encourages a pre-defined minimum number of voxels to predict each partition.
\item Representation learning using AutoAtlas' features that are uniquely associated with each partition. 
\end{itemize}

We evaluate the predictive performance of AutoAtlas' features for several measures included in the HCP-YA dataset \cite{Van_Essen2013-xf, Glasser2016-kr, Smith2013-vw, Barch2013-wy, Van_Essen2012-cf, glasser2013minimal} using several regressors, including linear regression, support vector machine, multilayer perceptron, and nearest neighbor methods.
We compare the performance using AutoAtlas features and features associated with established anatomical regions as produced by Freesurfer \cite{Desikan2006-zh} that is included in the HCP-YA dataset.
Using the prediction results, we then compute importance scores for each partition and map them back onto the input brain volumes, to show partitions that are more important for the given prediction task.
We have publicly released the code for AutoAtlas under an open-source license at \url{https://github.com/LLNL/autoatlas}.

\section{Related Work}
This work is related to several areas that intersect brain imaging and computer vision.

\paragraph*{Segmentation and Landmark Collocation}
Supervised techniques for semantic segmentation include fully Convolutional Neural Networks (CNNs) \cite{Long2015-vs}, and the U-net architecture, which was developed initially for medical image segmentation \cite{unetronn2015, Cicek2016-oo}.
Unsupervised techniques for image segmentation are much more difficult in part because of the lack of objective performance criteria.
An unsupervised, but not data-driven, approach for medical imagery (e.g. Computed Tomography or abdominal MRI) has been developed \cite{Aganj2018-fq}.
However, as opposed to this method, the collocation of landmarks in images within an unsupervised framework is of interest in the present work \cite{Tang2014-qu, Cho2015-nc, Novotny2017-qu, Kanazawa2016-qs, Thewlis2017-ws}.
In particular, our work is motivated by a novel approach for collocation of facial landmarks \cite{Zhang2018-ys}.
In that work, a dual landmark detection and representation learning approach was taken to localize the salient landmarks.
In our work, we adopt a similar concept for 3D volumes instead of 2D images, and broaden the landmarks to 3D regions in the volume.
This necessitates the development of a novel loss function to encourage well-behaved regions.
We refer to the resulting 3D regions as \emph{partitions} rather than segments (and the process as \emph{partitioning} rather than segmenting) since our application differs from classical problems of segmentation in computer vision.

\paragraph*{Anatomical Neuroimaging Based Parcellation}
A host of atlases have been produced that subdivide the brain into physiologically relevant regions \cite{Arslan2018-wg}.
The method for segmentation varies from registration onto a fixed reference brain to semi-supervised areal classifier \cite{Glasser2016-qg, Desikan2006-zh, Destrieux2010-ag}.
Many approaches for data-driven prediction of anatomical regions utilize datasets of manually labeled regions \cite{De_Brebisson2015-in, Fischl2002-rj, Chen2018-af, Patenaude2011-zk}.
Methodologies also exist for segmentation of brain images containing pathologies \cite{Havaei2017-fa, Mazzara2004-wx}.
Unsupervised approaches for brain segmentation include coarse white matter/gray matter segmentation \cite{Rivest-Henault2011-wy,Kong2015-te}.
Our work differs from previous approaches in that we do not attempt to recover anatomical regions, but instead simultaneously learn segmentation and latent representations for each partition in a purely data-driven approach.

\paragraph*{Prediction from MRI}
Machine learning methods for prediction of clinical diagnoses using FreeSurfer based features shows that the choice of machine learning approach may not be as important as the choice of input features \cite{Sabuncu2015-sp}.
An approach for creating saliency maps, the relevance voxel machine, overlaid on the MRI has been developed specifically targeting volumetric medical imagery \cite{Sabuncu2012-ti}.
Prediction of Alzheimer's Disease classification using a 3D CNN has been shown to out perform other techniques that utilize multiple imaging modalities \cite{Hosseini-Asl2016-bq}.
Prediction of subjects' age from structural MRI shows accuracy in the range of $\pm$10 years \cite{Franke2010-pe, Franke2012-cj}.
Our approach differs from these in that we first compute a spatially sensitive representation of the MRI data, followed by training a machine learning predictor.
This approach allows us to formulate a set of partitions of the brain by which we can compare results across individuals.


\section{Proposed Approach}

\begin{figure*}[!tb]
    \begin{center}
    \includegraphics[width=7in,keepaspectratio=true]{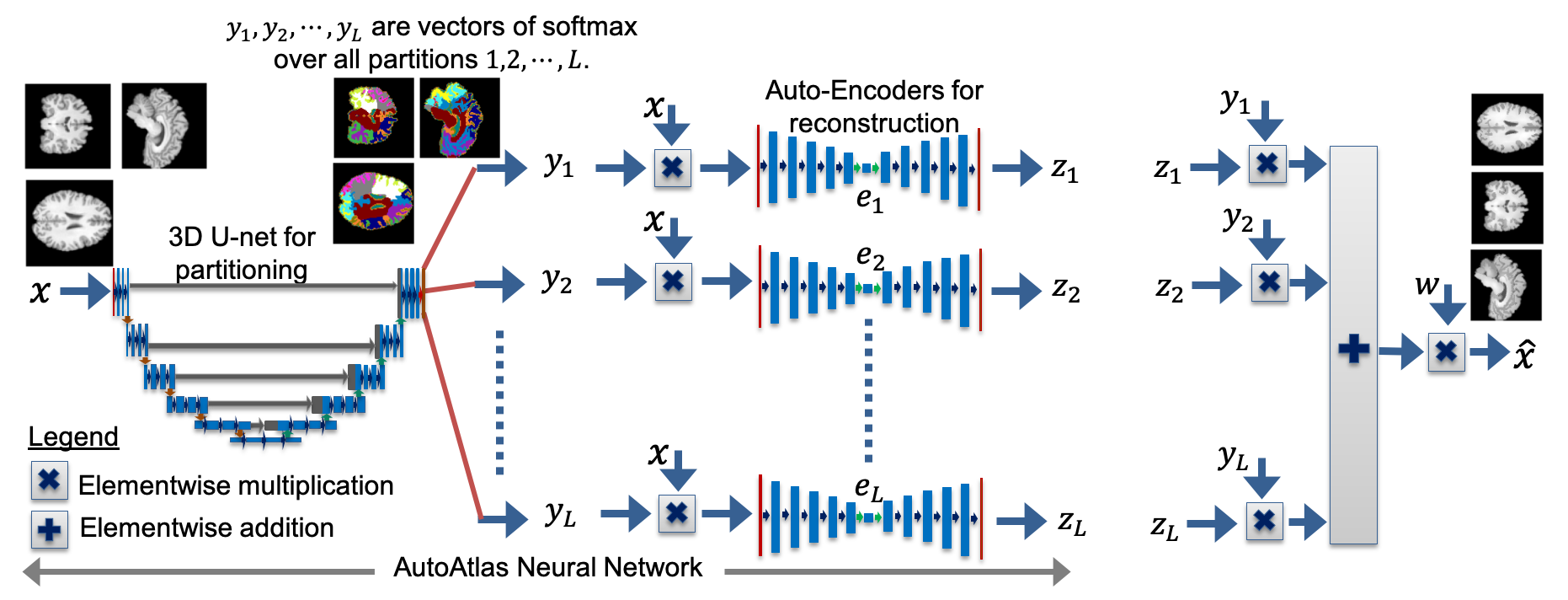}
    \end{center}
    \caption{\label{fig:fullblkdiag}
    AutoAtlas consists of a 3D CNN, such as the U-net in this paper, for partitioning the input volume into several partitions each of which is reconstructed using a low-capacity autoencoder. The U-Net is used for 3D multi-label partitioning of the input volume $x$. The $i^{th}$ autoencoder reconstructs the region within the $i^{th}$ partition. The flow chart to the right is a method to
reconstruct the volume $\hat{x}$ from autoencoder reconstructions $z_i$. However, the reconstructions $z_i$ are directly used in the loss functions (section \ref{sec:lossfuncs}) instead of $\hat{x}$. }
\end{figure*}

AutoAtlas is a novel neural network architecture for fully unsupervised partitioning
and representation learning of 3D volumes. It learns to partition the input into multiple regions by consistently recognizing and sparsely encoding the common morphological structure of each region across volumes. 
The partitions generated by AutoAtlas adapt to variations in 
morphology and texture from volume to volume. 

\subsection{Neural Network Architecture}
The architecture of AutoAtlas is shown in Fig. \ref{fig:fullblkdiag}. 
A detailed description of AutoAtlas' neural-network components is presented in Appendix \ref{app:nnarch}.
A 3D CNN  is used to partition the input volume into multiple partitions.
We use a 3D U-Net \cite{unetronn2015,3dunet2016} as the 3D CNN for partitioning the MRI brain scans from several subjects. 
In Fig. \ref{fig:fullblkdiag}, the input to the 3D U-Net is the 3D MRI brain volume, which is compactly represented in vector form as $x=[x_1,x_2,\cdots,x_N]$ where $x_j$ is the value at the $j^{th}$ input voxel and $N$ is the total number of voxels. 
The output of the 3D U-Net are vectors of classification probabilities $y_1,y_2,\cdots,y_L$, where $L$ is the total number of partitions and $y_i$ indicates the likelihood of voxels to belong to a particular partition $i$.
Note that $y_i=[y_{i,1},y_{i,2},\cdots,y_{i,N}]$, where $y_{i,j}$ is the probability that voxel $j$ belongs to partition $i$. 
The probabilities $y_{i,j}, 1\leq i\leq L$ at each voxel $j$ are obtained by a $L$ dimensional softmax at the output of the U-net.  

The U-net is followed by multiple low capacity autoencoders where
each autoencoder is tasked with the reconstruction of one partition.
The $i^{th}$ autoencoder from the top in Fig. \ref{fig:fullblkdiag} 
strives to accurately reconstruct the input MRI volume $x$
at its output $z_i$ within partition $i$ (where $y_{i,j} \approx 1$).
The input to the $i^{th}$ autoencoder is the elementwise vector product of the MRI volume $x$ 
and the partition probabilities $y_i$.
Each layer of the autoencoder has the same number of channels $C_a$. 
At the bottleneck layer where the contracting path ends and the expanding path begins, the total number of activations is also $C_a$.  
These activations, henceforth called feature embeddings,  represent an encoding of the input volume within the $i^{th}$ partition (where $y_{i,j} \approx 1$) and are useful for representation learning.

\subsection{Loss Functions}
\label{sec:lossfuncs}
To train AutoAtlas, we minimize the sum of three loss functions each of which serve a unique purpose. 
The first is Reconstruction Error (RE) loss that is used to measure the quality of 
autoencoder reconstruction for each partition by
comparing the voxel values within the corresponding partition in the input volume.
The second is the Neighborhood Label Similarity (NLS) loss that forces 
neighboring voxels to predict the same label with high confidence. 
Lastly, we have the Anti-Devouring (AD) loss that imposes a penalty
if the number of voxels that belong to a given partition is less than a pre-defined threshold.
The AD loss prevents partitions where a few labels devour all
the other labels and represent the whole volume.

\subsubsection{Reconstruction Error (RE) Loss}
The RE loss is used to ensure that the partitioning
is not arbitrary but rather conforming to the distinctive morphological structure in the input brain volumes. 
It drives the output of every $i^{th}$ autoencoder, $z_i$, to match the 
values in the input brain volume, $x$, within the $i^{th}$ partition.
The RE loss penalizes differences between the output of the autoencoder, $z_i$, and the input volume, $x$,
while weighting the penalty terms by the label probabilities $y_i$. 
The penalty terms are also weighted by the foreground mask vector $w$,
where $w_j=1$ for voxels within the brain and $w_j=0$ for voxels outside the brain.
The weight parameter $w_j$ ensures that AutoAtlas ignores the partitioning of background region
with no brain tissue.
Thus, the RE loss is given by,
\begin{equation}
\label{eq:relloss}
L_{RE} = \frac{1}{\bar{w}}\sum_{i=1}^{L}\sum_{j=1}^N w_j y_{i,j} \left|x_j-z_{i,j}\right|^2
\end{equation}
where  $\bar{w}$ is the total number of voxels within the foreground region given by $\bar{w}=\sum_{j=1}^N w_j$, 
$L$ is the total number of partitions, 
$x_j$ is the value of input volume $x$ at voxel index $j$, 
$y_{i,j}$ is the probability of the $i^{th}$ partition label at the $j^{th}$ voxel 
obtained using the softmax output of the U-net,
and $z_{i,j}$ is the $j^{th}$ voxel value at the output of the $i^{th}$ autoencoder 
that is focused on reconstruction in the $i^{th}$ partition.
Given the autoencoder outputs $z_{i,j}$, a reconstruction of voxel $j$ is given by,
\begin{equation}
\label{eq:mrirec}
\hat{x}_j = w_j \sum_{i=1}^{L} y_{i,j} z_{i,j}.
\end{equation}

\subsubsection{Neighborhood Label Similarity (NLS) Loss}
The NLS loss is used to 
encourage neighboring voxels to predict the same partition label with high confidence.
Two voxels are considered neighbors if they are adjacent in the sense that
there is a shared edge or vertex between the voxels.  
NLS loss not only encourages neighboring voxels to have similar classification probabilities
but also forces them to predict only one partition label with high probability.

To guide the formulation of NLS loss, 
we first define a measure of smoothness for AutoAtlas partitions given by, 
\begin{equation}
\label{eq:smoothmeas}
\sum_{i=1}^{L}\frac{1}{|\mathcal{N}|}\sum_{k,l\in \mathcal{N}} y_{i,k}y_{i,l},
\end{equation}
where $L$ is the total number of partitions,
$\mathcal{N}$ is the set of all pairwise indices of voxel neighbors within the foreground region 
of the brain where $w_j=1$,
and $|\mathcal{N}|$ is the cardinality of the set $\mathcal{N}$.
The term $\frac{1}{|\mathcal{N}|}\sum_{k,l\in \mathcal{N}} y_{i,k}y_{i,l}$ 
estimates the probability that any two neighboring voxels share the partition label $i$.
Hence, equation \eqref{eq:smoothmeas} estimates the                      
probability that any two neighboring voxels share the same (but any) partition label. 
Thus, increasing values for equation \eqref{eq:smoothmeas} improves the smoothness of the partitions
and also results in high confidence predictions by ensuring
\begin{equation}
\label{eq:confvoxs}
\max_i y_{i,j} \approx 1.
\end{equation} 
Suppose the entire brain is assigned a single partition label 
$\hat{i}$ i.e., $y_{\hat{i},j}\approx 1 \,\,\forall j$, then equation \eqref{eq:smoothmeas} will become one.  

Our design goal for the NLS loss function is to ensure that equation \eqref{eq:smoothmeas} is maximized.
Hence, NLS loss is defined as the negative logarithm of equation \eqref{eq:smoothmeas} and is given by, 
\begin{equation}
\label{eq:nsslloss}
L_{NLS} = -\log\left(\sum_{i=1}^{L}\frac{1}{|\mathcal{N}|}\sum_{k,l\in \mathcal{N}} y_{i,k}y_{i,l}\right)
\end{equation}
By minimizing the NLS loss, we encourage the resulting partitions to be smooth, localized, and contiguous. 

\subsubsection{Anti-Devouring (AD) Loss}
The AD loss is used to encourage every partition label to be associated
with at least a predefined minimum fraction of the total number of voxels in the volume.
In the absence of this loss, only a few labels would be chosen to partition the brain by the partitioning CNN, the U-net in Fig. \ref{fig:fullblkdiag}, which may lead to large partitions that cannot be accurately reconstructed by the low capacity autoencoders.
This in turn leads to insufficient reduction in RE loss and the training converges to a sub-optimal solution.
Using AD loss, we are able to drive the partitioning CNN to 
use all of the partition labels, thereby presenting each autoencoder with
a small sub-region of the brain for encoding and reconstruction.

Our design goal with AD loss is to ensure that every partition satisfies the AD constraint given by,
\begin{equation}
\label{eq:adconspart}
\frac{1}{\bar{w}}\sum_{j=1}^N w_j y_{i,j} \geq u_i,
\end{equation} 
where  $0<u_i\leq 1$ is the desired minimum frequency of occurrence for label $i$.
Note that a voxel $j$ is said to predict a partition label $\hat{i}$ when 
\begin{equation}
\label{eq:partpred}
\hat{i}=\arg\max_{i} y_{i,j}.
\end{equation}
In association with the NLS loss that forces $\max_{i} y_{i,j} \approx 1$, 
equation \eqref{eq:adconspart} ensures that each partition $i$
is predicted by at least $u_i$ fraction of voxels within the brain 
($w_j=1$ defines the interior of the brain). 

To ensure that the partitions satisfy equation \eqref{eq:adconspart}, 
we minimize the following AD loss,
\begin{equation}
\label{eq:adlloss}
L_{AD} =  \frac{1}{L}\sum_{i=1}^{L}\max\left\lbrace -\log\left(\frac{1}{u_i\bar{w}}\sum_{j=1}^N w_j y_{i,j} + \epsilon\right),0 \right\rbrace,
\end{equation}
where $\epsilon$ is a very small value that provides numerical stability.
Conversely, if equation \eqref{eq:adconspart} is satisfied for $i$, 
then the contribution of partition $i$ to the 
AD loss in equation \eqref{eq:adlloss} is zero.
In equation \eqref{eq:adlloss}, the term $\frac{1}{\bar{w}}\sum_{j=1}^N w_j y_{i,j}$ 
is the average of softmax activations belonging to label $i$ over all voxels 
within the foreground region where $w_j=1$. 
If this term is greater than $u_i$, 
then it means that label $i$ has met the desired criteria
and does not contribute to the loss.
Alternatively, if this term is lower than $u_i$, then 
it means label $i$ is under-represented by voxels resulting
in positive contribution to the AD loss.
The minimum frequency parameter $u_i$ for AD loss is chosen such 
that it is inversely proportional to the number of labels, $L$. 
In particular, we choose $u_i=\frac{c}{L}$,
where $c$ is a constant that is close to but less than one. 
We choose $\epsilon=10^{-10}$, which ensures that the logarithm 
in equation \eqref{eq:adlloss} is always defined even when $y_{i,j}=0$ for all $j$.

\subsubsection{Total Loss}
The total loss that is minimized during training is given by the weighted sum 
of RE, NLS, and AD losses,
\begin{equation}
\label{eq:totloss}
L_{TOT} = \lambda_{RE} L_{RE} + \lambda_{NLS}L_{NLS} + \lambda_{AD} L_{AD},
\end{equation}
where $\lambda_{RE}$, $\lambda_{NLS}$, and $\lambda_{AD}$ are the regularization weight hyper-parameters for RE loss, NLS loss, and AD loss respectively.
In this paper, $\lambda_{RE}$ is always chosen to be equal to $1$ unless specified otherwise. 

\subsection{Representation Learning}
\label{sec:replearn}
In the HCP-YA \cite{Van_Essen2013-xf, Glasser2016-kr, Smith2013-vw, Barch2013-wy, Van_Essen2012-cf} dataset, in addition to the MRI scans, several functional
and behavioral meta data were collected for each individual subject.
AutoAtlas produces several low-dimensional embedding features summarizing
each MRI brain volume that are useful to predict such meta data.
The dimensionality of the embedding features $e_i$ at the bottleneck layer of each 
autoencoder is $C_a$, which in our application is chosen to be either $4$, $8$, $16$, $32$, or $64$.
Embedding $e_i$ is an encoding of the input volume $x$ multiplied by the classification probabilities $y_i$ as shown in Fig. \ref{fig:AutoEnc}.
Importantly, $e_i$ is a $C_a$ dimensional representational encoding of the $i^{th}$ partition. 
For representation learning, we combine the embeddings from all partitions as, 
\begin{equation}
\label{eq:ftall}
E = \left[e_1, e_2, \cdots, e_L\right]. 
\end{equation}
The total dimensionality of embeddings in $E$ that includes $e_i$ from all the autoencoders is $LC_a$, where $L$ is the total number of partitions.
Thus, $E$ represents a low-dimensional encoding of the MRI brain volume that is useful to predict other meta data in the HCP-YA dataset.

\vspace{-0.1in}
\subsection{Data and Hyperparameters}
\begin{figure}[!tb]
    \begin{center}
    \includegraphics[width=3.4in,keepaspectratio=true]{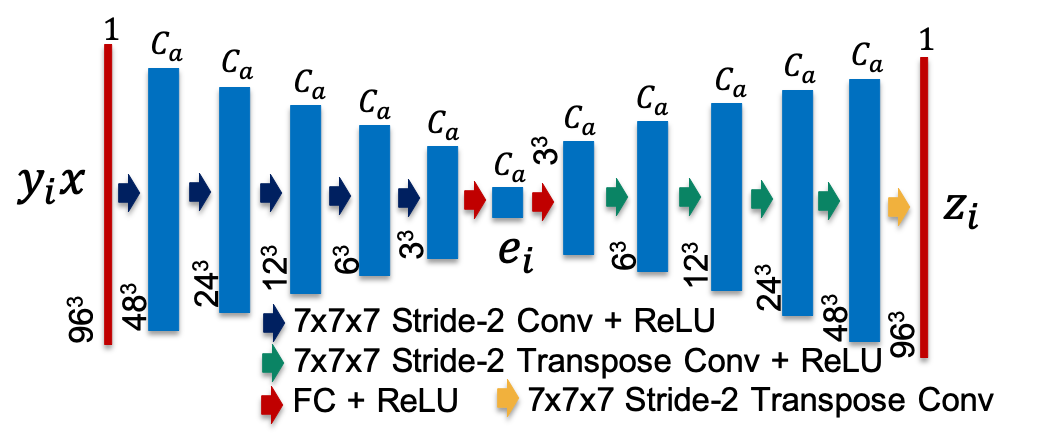}
    \end{center}
    \caption{\label{fig:AutoEnc}
    3D autoencoder for reconstruction of voxels contained within partition $i$. 
    $e_i$ is a vector denoting the embedding features in the bottleneck layer.
	Each rectangle represents a tensor. The number of channels is indicated above the tensor. 
 The spatial size is indicated to the left of the tensor.}
\end{figure}

We use the Adam optimizer \cite{kingma2014adam} to minimize the total loss function given
 in equation \eqref{eq:totloss} by jointly
optimizing the parameters of the 3D U-net and the autoencoders.
For the Adam optimizer, we use a fixed learning rate of $0.0001$, momentum parameter of $0.9$ for the running average of the gradient, and momentum parameter of $0.999$ for the running average of the square of gradient.
We use $C_s=32$ number of channels at the first layer output of U-net shown in Fig. \ref{fig:UNet}.
Each input volume is partitioned into $L=16$ partitions
and hence the softmax at the output layer of the U-net in Fig. \ref{fig:UNet} is 
over $16$ channels.
Our choice of $L=16$ was driven by the limit on the maximum available GPU memory.
Larger values for $L$ encourage smaller sized partitions
with improved localization and more efficient representation
 in the feature embeddings, $e_i$, of autoencoders.
The number of channels, $C_a$, for each autoencoder in Fig. \ref{fig:AutoEnc} 
is chosen to be either $4$, $8$, $16$, $32$, or $64$.

AutoAtlas was trained on a total of $859$ MRI
volumes for $500$ epochs
and tested on $215$ volumes from the HCP-YA dataset \cite{Van_Essen2013-xf, Glasser2016-kr, Smith2013-vw, Barch2013-wy, Van_Essen2012-cf, glasser2013minimal}.
All original MRI volumes were down-sampled to a voxel resolution of $2mm$ and zero-padded so that the final size of each volume was $96\times 96 \times 96$.
Next, all volumes are normalized by dividing 
by the average of the maximum voxel value computed for each volume in the training set.
AutoAtlas was then trained using  a batch size of $4$ on four NVIDIA Tesla V100 GPUs. 
We used PyTorch \cite{paszke2017automatic} machine learning framework for implementing AutoAtlas. 
We used the same train-test split when predicting meta-data in the HCP-YA dataset 
from AutoAtlas' embedding features. 
After excluding samples with missing meta-data, 
we had $856$ samples for training and $215$ samples for testing.

Unless specified otherwise, we always use $\lambda_{RE}=1$ for the RE loss. While it is possible to get consistent partitions for 
a wide range of values for the regularization parameters $\lambda_{NLS}$ and $\lambda_{AD}$, 
incorrect values can still lead to sub-optimal partitions 
if either of these two regularization parameters are too small. 
For instance, if $\lambda_{NLS}=0$, then the U-net is not required by design to produce smooth and contiguous partitions. 
Alternatively, if $\lambda_{AD}=0$, then the U-net is not required to
use all the available labels and hence it chooses only a small number 
of labels to partition the whole volume. 
In this case, due to the large size and complex morphology of each partition, 
the autoencoder will not be able to encode and reconstruct it.
Thus, consistent partitions are achieved only when both
the regularization parameters are set to a sufficiently large value.
In our application, we used $\lambda_{RE}=1$, $\lambda_{NLS}=0.005$, $\lambda_{AD}=0.1$, and $u_i=\frac{0.9}{L}=0.05625$
to train all the AutoAtlas models.
The foreground mask $w$ in equation \eqref{eq:relloss} is computed by binary thresholding
each MRI volume in the training and testing sets.

\makeatletter
\define@key{Gin}{segsz}[true]{%
    \edef\@tempa{{Gin}{width=1.0in,keepaspectratio=true}}%
    \expandafter\setkeys\@tempa
}
\makeatother

\begin{figure}[!htb]
\begin{center}
Orthogonal slices of MRI volume  \\
\begin{tabular}{ccc}
\includegraphics[segsz]{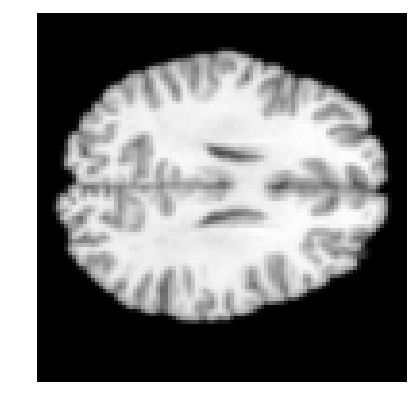} & 
\includegraphics[segsz]{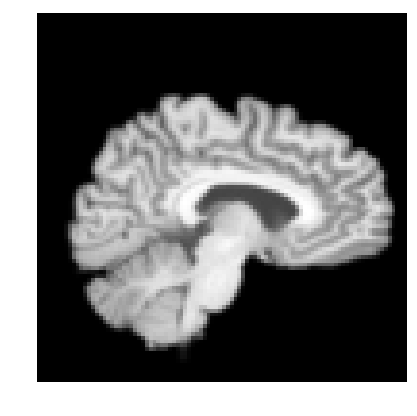} &
\includegraphics[segsz]{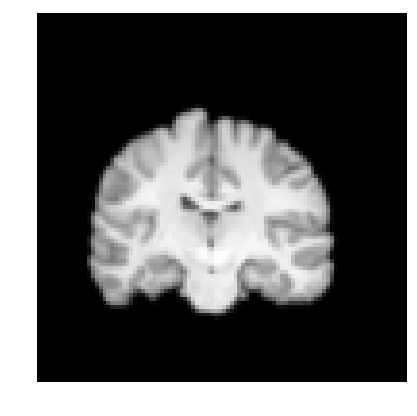} \vspace{-0.1in}\\
(a) & (b) & (c) \\
\end{tabular}\\
\vspace{0.03in}
Corresponding tissue segmentations \\
\begin{tabular}{ccc}      
\includegraphics[segsz]{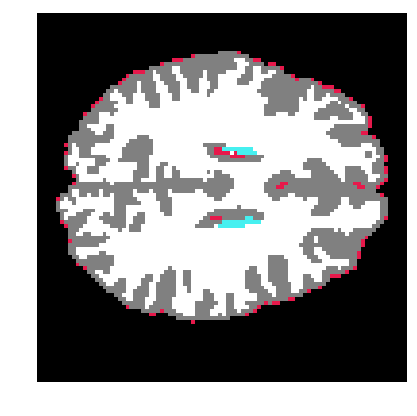} & 
\includegraphics[segsz]{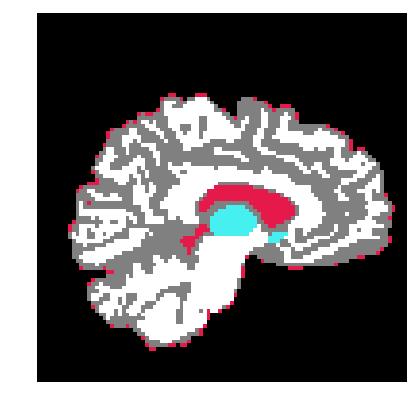} &
\includegraphics[segsz]{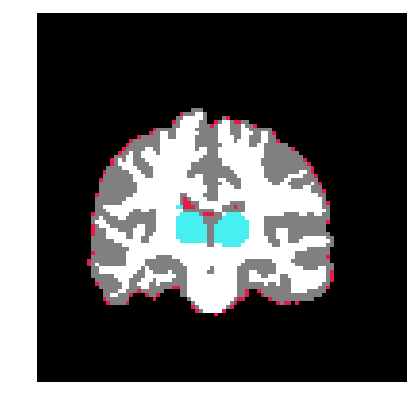} \vspace{-0.1in}\\
(d) & (e) & (f) \\
\end{tabular}\\
\vspace{0.03in}
AutoAtlas partitions with $C_a=16$\\
\begin{tabular}{ccc}      
\includegraphics[segsz]{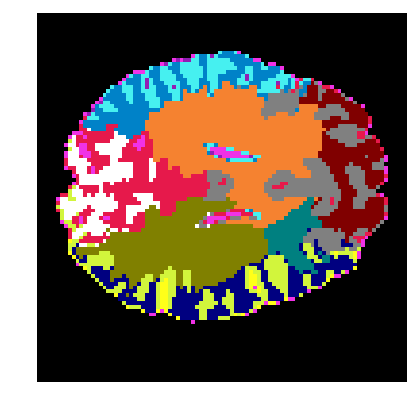} & 
\includegraphics[segsz]{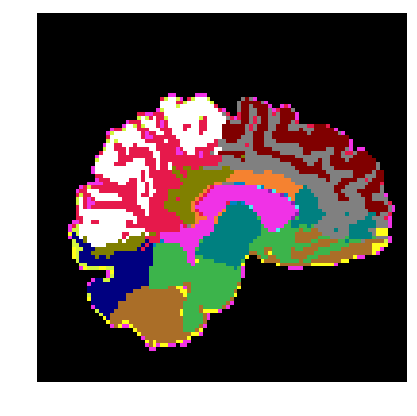} &
\includegraphics[segsz]{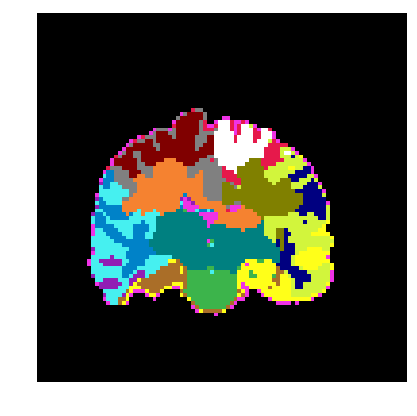} \vspace{-0.1in}\\
(g) & (h) & (i) \\
\end{tabular}\\
\end{center}
\vspace{-0.1in}
\caption{\label{fig:gtseg_zyx}
First row shows MRI image slices along three mutually orthogonal planes for the same subject. 
Second row shows the corresponding tissue segmentation for images in the first row.
Third row shows the resulting partitions using AutoAtlas with $C_a=16$.
}
\end{figure}

\begin{figure}[!htb]
\begin{center}
\begin{tabular}{cccc}
\includegraphics[width=0.7in]{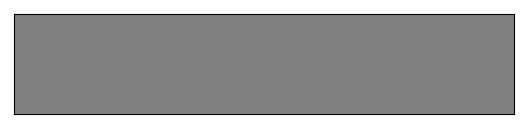} & 
\includegraphics[width=0.7in]{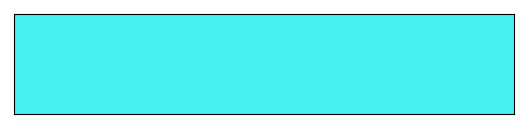} &
\includegraphics[width=0.7in]{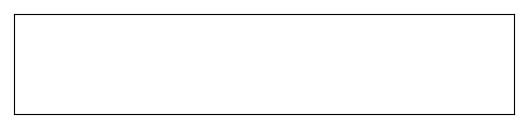} &
\includegraphics[width=0.7in]{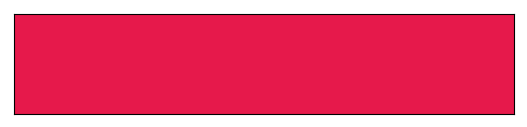} \vspace{-0.05in}\\
Label 0 & Label 1 & Label 2 & Label 3 \\
\vspace{0.02in}
\includegraphics[width=0.7in]{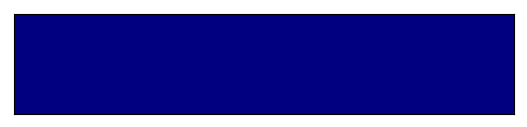} & 
\includegraphics[width=0.7in]{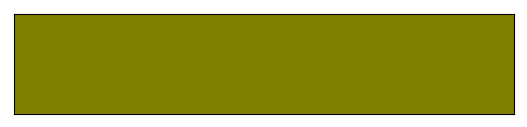} &
\includegraphics[width=0.7in]{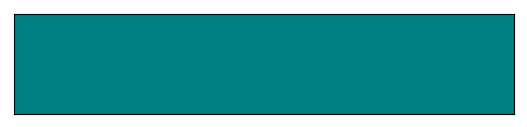} &
\includegraphics[width=0.7in]{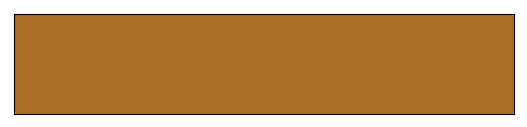} \vspace{-0.05in}\\
Label 4 & Label 5 & Label 6 & Label 7 \\
\vspace{0.02in}
\includegraphics[width=0.7in]{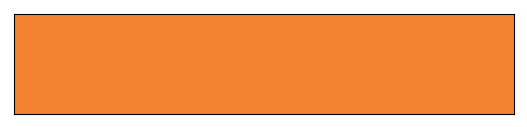} & 
\includegraphics[width=0.7in]{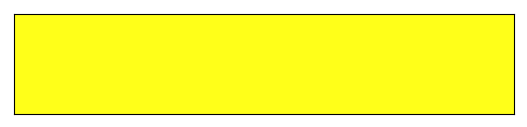} &
\includegraphics[width=0.7in]{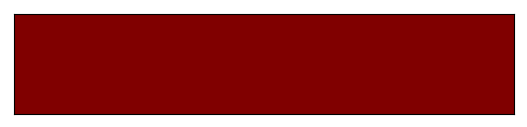} &
\includegraphics[width=0.7in]{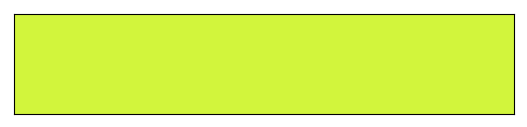} \vspace{-0.05in}\\
Label 8 & Label 9 & Label 10 & Label 11 \\
\vspace{0.02in}
\includegraphics[width=0.7in]{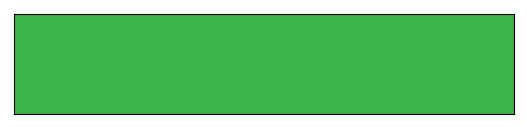} & 
\includegraphics[width=0.7in]{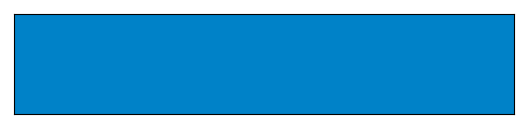} &
\includegraphics[width=0.7in]{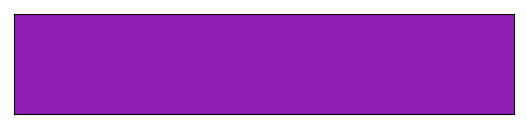} &
\includegraphics[width=0.7in]{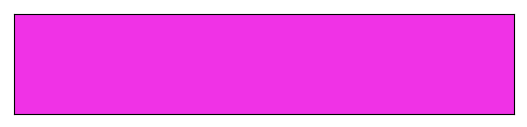} \vspace{-0.05in}\\
Label 12 & Label 13 & Label 14 & Label 15 \\
\end{tabular}
\end{center}
\vspace{-0.1in}
\caption{\label{fig:colleg} Legend indicating the color associated with each AutoAtlas label.}
\end{figure}

\begin{figure*}[!htb]
\begin{center}
Image slices through the center of MRI brain volumes of multiple subjects\\
\begin{tabular}{cccccc}
\includegraphics[segsz]{figs/v2/emb16/100206/T1_z.png} & 
\includegraphics[segsz]{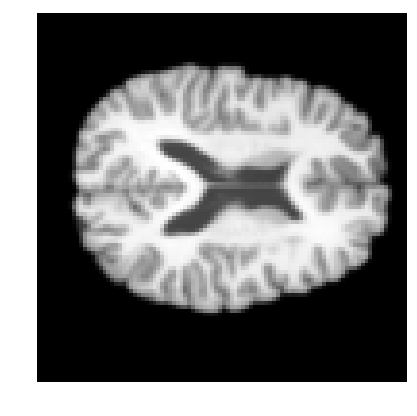} &
\includegraphics[segsz]{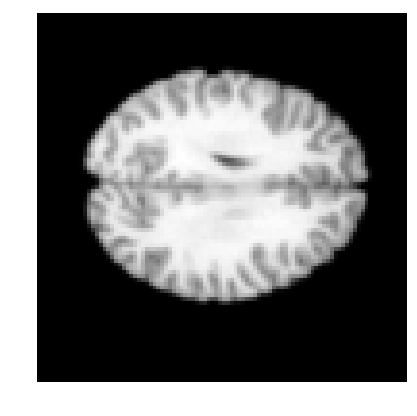} &
\includegraphics[segsz]{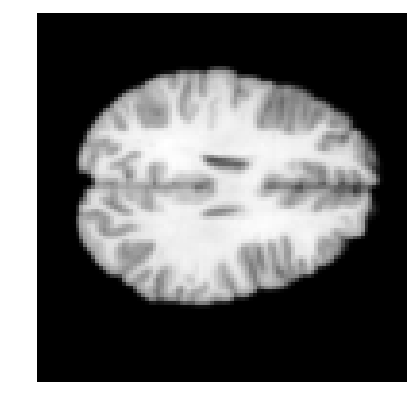} & 
\includegraphics[segsz]{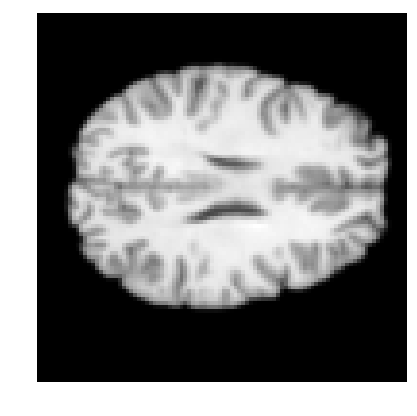} &
\includegraphics[segsz]{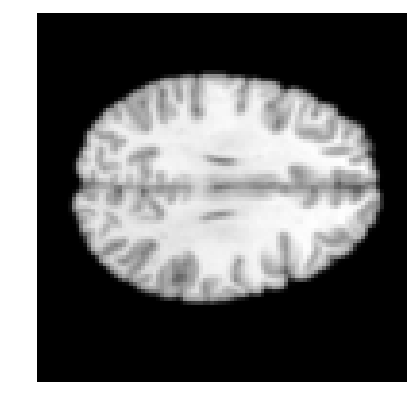} \vspace{-0.1in}\\
(a) & (b) & (c) & (d) & (e) & (f) \vspace{0.05in}\\
\end{tabular}\\
Tissue segmentations where each color represents a unique tissue type\\
\begin{tabular}{cccccc}      
\includegraphics[segsz]{figs/v2/emb16/100206/tt5_z.png} & 
\includegraphics[segsz]{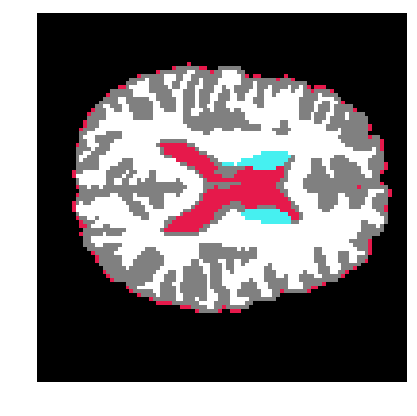} &
\includegraphics[segsz]{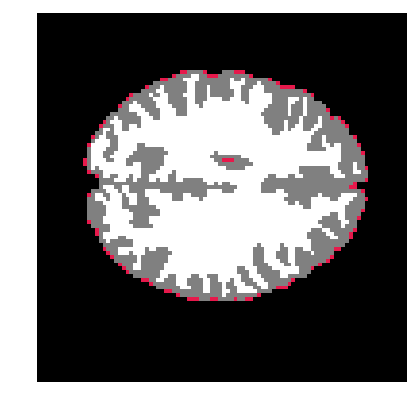} &
\includegraphics[segsz]{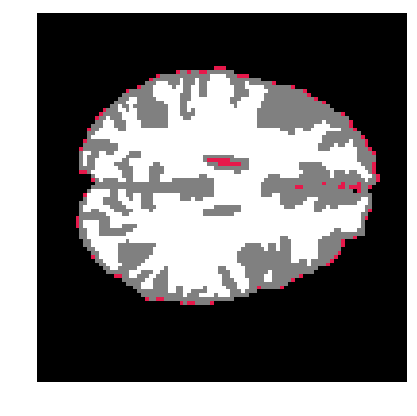} & 
\includegraphics[segsz]{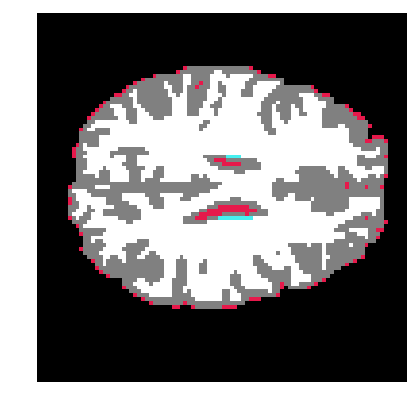} &
\includegraphics[segsz]{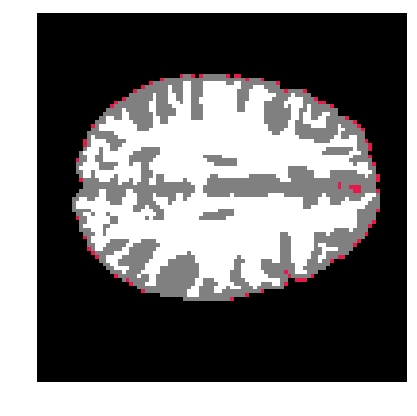} \vspace{-0.1in}\\
(g) & (h) & (i) & (j) & (k) & (l)\vspace{0.05in} \\
\end{tabular}\\
AutoAtlas partitioning with $C_a=16$ where each color represents a unique partition\\
\begin{tabular}{cccccc}      
\includegraphics[segsz]{figs/v2/emb16/100206/seg_z.png} & 
\includegraphics[segsz]{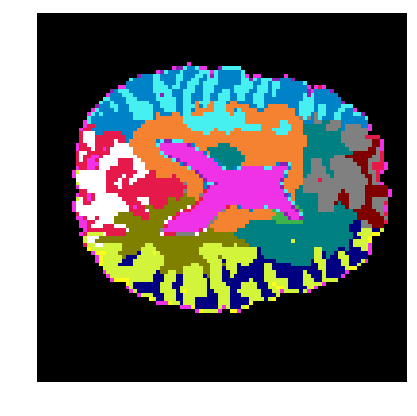} &
\includegraphics[segsz]{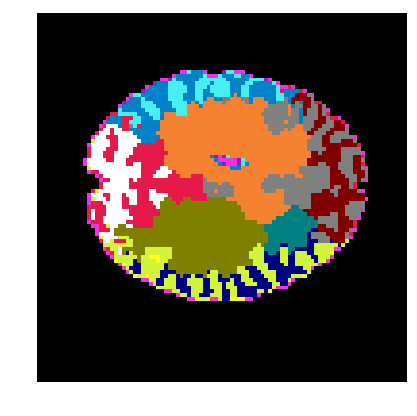} &
\includegraphics[segsz]{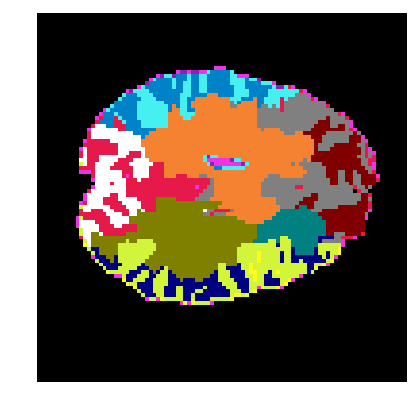} & 
\includegraphics[segsz]{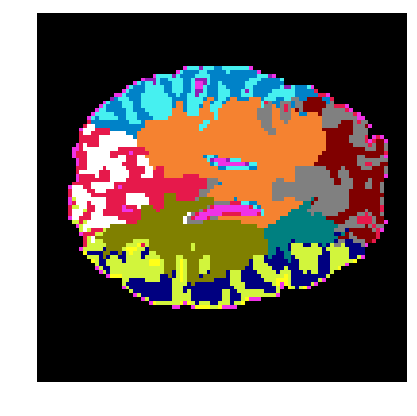} &
\includegraphics[segsz]{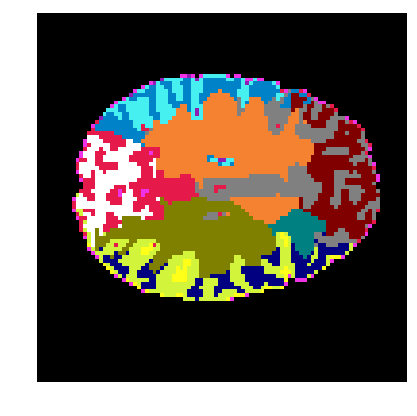} \vspace{-0.1in}\\
(m) & (n) & (o) & (p) & (q) & (r)\vspace{0.05in} \\
\end{tabular}\\
\end{center}
\vspace{-0.15in}
\caption{\label{fig:gtseg_subj}
Comparison of AutoAtlas partitions with tissue segmentations across subjects. 
First row shows image slices through the center of MRI brain volumes across multiple subjects. 
Second row shows segmentations of the brain into four different tissue types. 
Third row shows the partitions generated using AutoAtlas with $C_a=16$. 
The images along each column are for the same subject, while the images along a row are for different subjects. 
The partitioning of AutoAtlas adapt to variations in structure of brain tissue while consistently identifying
the same region across subjects. 
}
\end{figure*}

\begin{table}[!htb]
\caption{\label{tab:olapaatt} 
Overlap between AutoAtlas ($AA$) partitions and regions representing various tissue types ($TT$).
The overlap percentage between the $i^{th}$ AutoAtlas partition (indicated by $AA_i$) and $j^{th}$ tissue region (indicated by $TT_j$) is computed as the number of voxels that lie in the intersection of the two ($AA_i$ and $TT_j$) expressed as a percentage of the total number of voxels that lie inside the brain. The mean (first value) and standard deviation (second value within brackets) of the overlap percentage across subjects is shown for every pair of AutoAtlas partition and tissue region.
Bold font is used to highlight the fields for which the mean overlap percentage is greater than one.
For the tissue types, $TT_0$ is gray matter, $TT_1$ is sub-cortical, $TT_2$ is white matter, and $TT_3$ is fluid.
AutoAtlas used an embedding size of $C_a=16$.}
\begin{center}
\begin{tabular}{|c|c|c|c|c|} 
\hline
& $TT_0$ & $TT_1$ & $TT_2$ & $TT_3$ \\\hline
$AA_{0}$ & \textbf{5.95 (0.20)} & 0.00 (0.00) & 0.27 (0.04) & 0.06 (0.04)\\\hline
$AA_{1}$ & \textbf{5.34 (0.20)} & 0.02 (0.01) & 0.76 (0.10) & 0.00 (0.00)\\\hline
$AA_{2}$ & \textbf{1.36 (0.16)} & 0.00 (0.00) & \textbf{5.01 (0.23)} & 0.00 (0.00)\\\hline
$AA_{3}$ & \textbf{5.61 (0.18)} & 0.00 (0.00) & 0.29 (0.08) & 0.18 (0.09)\\\hline
$AA_{4}$ & \textbf{3.27 (0.31)} & 0.00 (0.00) & \textbf{3.41 (0.34)} & 0.00 (0.00)\\\hline
$AA_{5}$ & 0.13 (0.04) & 0.00 (0.00) & \textbf{6.04 (0.23)} & 0.00 (0.00)\\\hline
$AA_{6}$ & 0.20 (0.04) & \textbf{1.33 (0.08)} & \textbf{4.61 (0.19)} & 0.00 (0.00)\\\hline
$AA_{7}$ & \textbf{4.08 (0.32)} & 0.00 (0.00) & \textbf{2.19 (0.36)} & 0.00 (0.00)\\\hline
$AA_{8}$ & 0.18 (0.03) & 0.37 (0.04) & \textbf{5.41 (0.15)} & 0.00 (0.00)\\\hline
$AA_{9}$ & \textbf{5.22 (0.20)} & 0.07 (0.02) & 0.67 (0.12) & 0.23 (0.05)\\\hline
$AA_{10}$ & \textbf{2.16 (0.22)} & 0.00 (0.00) & \textbf{4.02 (0.23)} & 0.00 (0.00)\\\hline
$AA_{11}$ & \textbf{5.80 (0.18)} & 0.00 (0.00) & 0.81 (0.10) & 0.00 (0.00)\\\hline
$AA_{12}$ & \textbf{1.65 (0.12)} & 0.54 (0.06) & \textbf{4.07 (0.21)} & 0.00 (0.00)\\\hline
$AA_{13}$ & \textbf{1.51 (0.13)} & 0.01 (0.01) & \textbf{4.88 (0.22)} & 0.00 (0.00)\\\hline
$AA_{14}$ & \textbf{4.70 (0.47)} & 0.00 (0.00) & \textbf{1.51 (0.47)} & 0.05 (0.02)\\\hline
$AA_{15}$ & \textbf{1.90 (0.28)} & 0.01 (0.01) & 0.00 (0.01) & \textbf{4.11 (0.48)}\\\hline
\end{tabular}
\end{center}
\end{table}

\begin{figure}[!hbt]
    \begin{center}
    AutoAtlas with $\lambda_{RE}=1$,  $\lambda_{NLS}=0$, and $\lambda_{AD}=0.1$\\
    \begin{tabular}{ccc}    
    \hspace{-0.2in} 
    \includegraphics[width=1in,keepaspectratio=true]{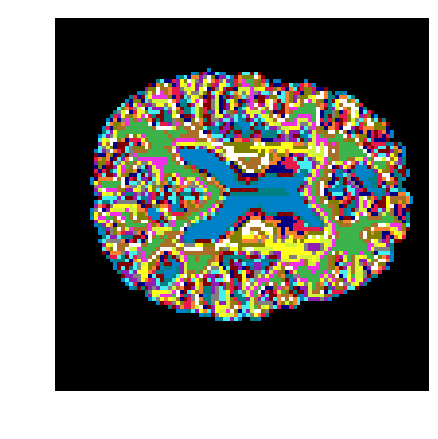} &   
    \hspace{-0.15in} 
     \includegraphics[width=1.22in,keepaspectratio=true]{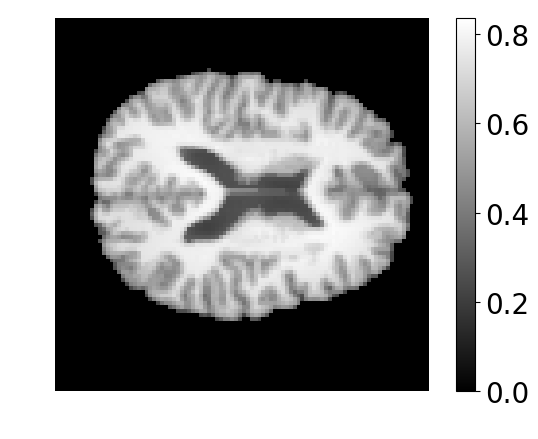} &
     \hspace{-0.22in}
      \includegraphics[width=1.22in,keepaspectratio=true]{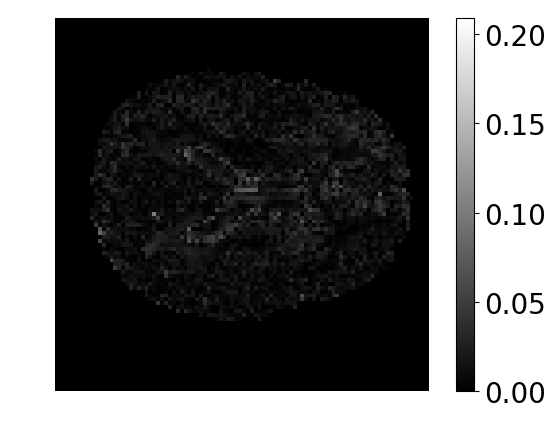} \vspace{-0.1in}\\ 
 \hspace{-0.2in} (a) Partitions &  \hspace{-0.15in}  (b) Reconstruction & \hspace{-0.22in} (c) Error\\
    \end{tabular}\\
    \vspace{0.05in}
    AutoAtlas with $\lambda_{RE}=1$,  $\lambda_{NLS}=0.005$, and $\lambda_{AD}=0$\\
    \begin{tabular}{ccc}
    \hspace{-0.2in} 
    \includegraphics[width=1in,keepaspectratio=true]{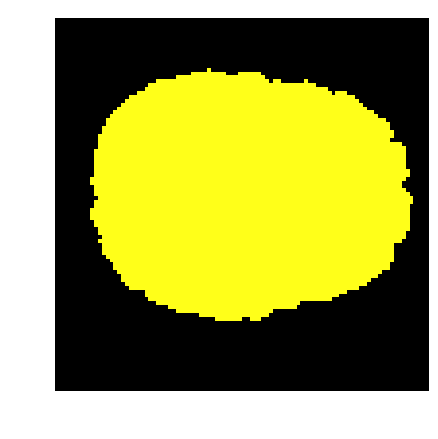} &   
    \hspace{-0.15in} 
     \includegraphics[width=1.22in,keepaspectratio=true]{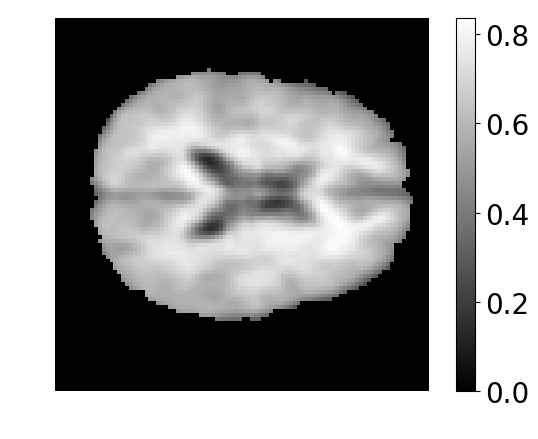} &
     \hspace{-0.22in}
      \includegraphics[width=1.22in,keepaspectratio=true]{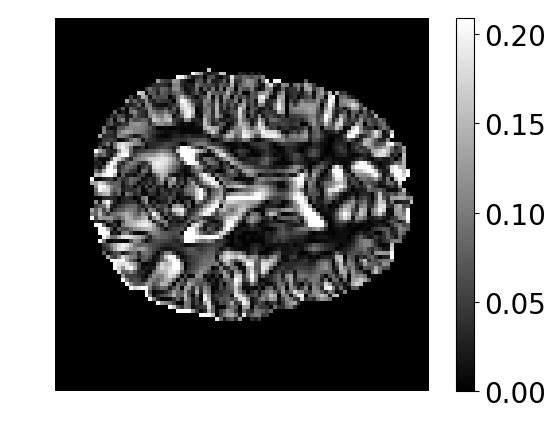} \vspace{-0.1in}\\ 
 \hspace{-0.2in} (d) Partitions &  \hspace{-0.15in}  (e) Reconstruction & \hspace{-0.22in} (f) Error\\
    \end{tabular}\\
    \vspace{0.05in}
    AutoAtlas with $\lambda_{RE}=1$,  $\lambda_{NLS}=0.005$, and $\lambda_{AD}=0.1$\\
    \begin{tabular}{ccc}
    \hspace{-0.2in} 
    \includegraphics[width=1in,keepaspectratio=true]{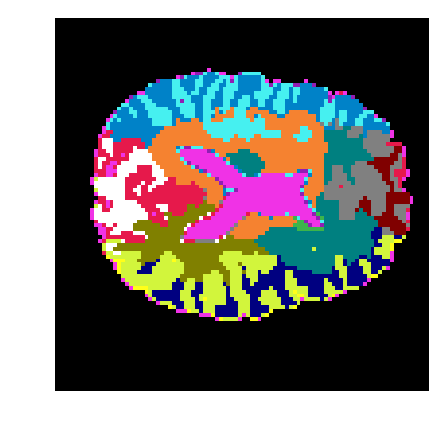} &   
    \hspace{-0.15in} 
     \includegraphics[width=1.22in,keepaspectratio=true]{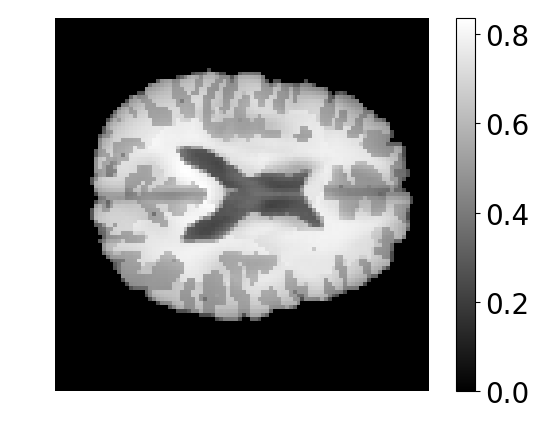} &
     \hspace{-0.22in}
      \includegraphics[width=1.22in,keepaspectratio=true]{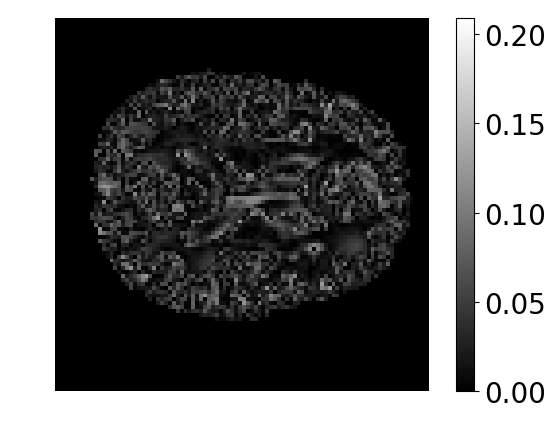} \vspace{-0.1in}\\ 
 \hspace{-0.2in} (g) Partitions &  \hspace{-0.15in}  (h) Reconstruction & \hspace{-0.22in} (i) Error\\
       \end{tabular}
    \end{center}
    \caption{\label{fig:ablsegc}
Ablation study of the NLS and AD losses (equations \eqref{eq:nsslloss} and \eqref{eq:adlloss}) 
by analyzing the partitioning and reconstruction performance of AutoAtlas.
Each color in (a,d,g) represents a unique partition. 
The reconstruction in (b,e,h) is computed using the flow-chart 
shown on the right side of Fig. \ref{fig:fullblkdiag}.
The reconstruction error in (c,f,i) is computed as the absolute value of the 
difference between the reconstruction and the ground-truth MRI.
By comparing (a) and (g), we see that the NLS loss is responsible for producing smooth and contiguous partitions. 
From (e,h) and (f,i), we see that incorporation of the AD loss produces a more accurate reconstruction by encouraging multiple partitions that each have their own representation.
 }
\end{figure}

\begin{figure}[!tb]
    \begin{center}
    \includegraphics[width=3.4in,keepaspectratio=true]{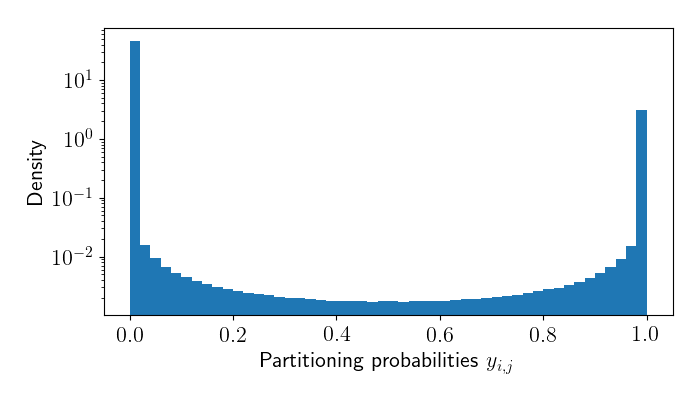}
    \end{center}
    \vspace{-0.3in}
    \caption{\label{fig:pphist}
   50 bin normalized histogram of the partitioning probabilities $y_{i,j}$ for all the volumes in the test set. Note that the $y-$axis is logarithm scaled. We can see that $y_{i,j}$ is predominantly either $0$ or $1$.}
\end{figure}

\section{Results}
In this section, we demonstrate the ability of AutoAtlas to perform
3D unsupervised partitioning of input volume into multiple partitions 
while producing low dimensional embedding features for each partition
that are useful for meta-data prediction.
Fig. \ref{fig:gtseg_zyx} is a comparison between various tissue types in an MRI volume and 
the partitions using AutoAtlas.
The white, gray, red, and blue colors in the tissue segmentation shown in Fig. \ref{fig:gtseg_zyx} (d-f)
represent the white matter, gray matter, fluid, and sub-cortical regions respectively.
These segmentations were produced using the FAST algorithm \cite{906424}.
The AutoAtlas labels associated with each color of the resulting partitions 
in Fig. \ref{fig:gtseg_zyx} (g-i) is shown in Fig. \ref{fig:colleg}. 
By comparing the AutoAtlas partitions in Fig. \ref{fig:gtseg_zyx} (g-i) 
with the tissue regions in Fig. \ref{fig:gtseg_zyx} (d-f),
we can see that the boundaries of AutoAtlas partitions conform to the tissue boundaries.
Fig. \ref{fig:gtseg_subj} demonstrates the consistency of AutoAtlas partitions across various subjects.
Each distinct color marks a unique partition that is consistently
recognized in the brain scans across multiple subjects. 
Importantly, the structure of each partition varies to account 
for the morphological changes in brain tissue across subjects. 
The autoencoder plays an important role to achieve this behavior by
sparsely encoding and decoding the partition associated with each label.
The images in Fig. \ref{fig:gtseg_zyx} and Fig. \ref{fig:gtseg_subj}
are for subjects from the test set using an AutoAtlas model with $C_a=16$.

Table \ref{tab:olapaatt} quantifies the consistency of brain partitioning across various subjects
by presenting the mean and standard deviation of the percentage of overlap 
between AutoAtlas partitions and various tissue types.
Each row in Table \ref{tab:olapaatt} corresponds to one AutoAtlas partition
and each column corresponds to a tissue type (white matter, sub-cortical region, gray matter, or fluid). 
The amount of overlap between AutoAtlas partitions and tissue types 
is quantified as a percentage of the total brain volume\footnote{Total brain volume is the number of voxels that contain brain tissue and excludes the background voxels.}.
The mean values of overlap percentage in Table \ref{tab:olapaatt}
indicate that each AutoAtlas partition overlaps significantly
with only one or two tissue types.
When compared to the mean, 
the relatively low values for standard deviation of the overlap percentage, 
shown in Table \ref{tab:olapaatt},
demonstrate that the percentage overlap is similar valued across subjects.
This demonstrates that each partition produced by AutoAtlas 
consistently learns to recognize and encode the same brain tissue.

The partitions in Fig. \ref{fig:gtseg_zyx} and Fig. \ref{fig:gtseg_subj} are 
a direct result of the interplay between the loss functions 
in equations \eqref{eq:relloss}, \eqref{eq:nsslloss}, and \eqref{eq:adlloss}.
An ablation study demonstrating the utility of the NLS loss and 
AD loss in AutoAtlas partitioning is shown in Fig. \ref{fig:ablsegc}.
The results in Fig. \ref{fig:ablsegc} and Fig. \ref{fig:gtseg_subj} (b,h,n) 
are for the same subject.
The reconstructions in Fig. \ref{fig:ablsegc} (b,e,h) are obtained by computing $\hat{x}=w\sum_{i=1}^L z_iy_i$
as shown in Fig. \ref{fig:fullblkdiag}. 
The reconstruction error in Fig. \ref{fig:ablsegc} (c,f,i)
is computed as $\left|\hat{x}-x\right|$, where $x$ is the ground-truth MRI that is input to AutoAtlas.
While Fig. \ref{fig:ablsegc} (a-c) and Fig. \ref{fig:ablsegc} (d-f) 
are for two different AutoAtlas models trained in the absence of the NLS loss ($\lambda_{NLS}=0$) 
and the AD loss ($\lambda_{AD}=0$) respectively, 
Fig. \ref{fig:ablsegc} (g-i) is for an AutoAtlas model that used all the three losses.
By comparing Fig. \ref{fig:ablsegc} (a-c) with Fig. \ref{fig:ablsegc} (g-i),
we can see that the NLS loss is necessary to get smooth and contiguous partitions
while still being able to accurately reconstruct the input MRI volume. 
By comparing Fig. \ref{fig:ablsegc} (d-f) with Fig. \ref{fig:ablsegc} (g-i),
we can see that the AD loss contributes to efficient partitioning of input volume into 
several representable partitions that are accurately reconstructed by the autoencoders.
Note that $y_{i,j}\approx 1$ if voxel $j$ is within the $i^{th}$ partition 
and $y_{i,j}\approx 0$ otherwise (Fig. \ref{fig:pphist}). 
The purpose of the mask $w$ is to ignore the partitioning and reconstruction in the background 
where there is no brain tissue. 

For representation learning, we compare prediction performances obtained by feeding
the embedding features, $E$ in equation \eqref{eq:ftall}, as input to machine learners such as ridge regression, nearest neighbor regression,
support vector machine (SVM), and multi-layer perceptron (MLP). 
As a baseline, we also report performance scores for predictors trained
 using $201$ structural features extracted by Freesurfer \cite{Desikan2006-zh}. 
We evaluated prediction of all meta-data related to motor skills
with the names \textit{Strength-Unadj}, \textit{Strength-AgeAdj}, \textit{Endurance-Unadj}, 
\textit{Endurance-AgeAdj}, \textit{Dexterity-Unadj}, \textit{Dexterity-AgeAdj}, and \textit{GaitSpeed-Comp}. 
However, we noticed that both Freesurfer features and AutoAtlas features
were only predictive for the meta-data, \textit{Strength-Unadj} and \textit{Strength-AgeAdj}. 
\textit{Strength-Unadj} is a measure of the grip strength of each hand and 
\textit{Strength-AgeAdj} measures grip strength while correcting for variations due to age.

\begin{table*}[h!]
\caption{\label{tab:predscores} Performance comparison between AutoAtlas and Freesurfer feature based regressors.
Legend: FS - Freesurfer features, $AAK$ - AutoAtlas features with $C_a=K$ embedding size for each autoencoder, Lin - Ridge Regression, NNbor -  Nearest Neighbor, SVM - Support Vector Machine, MLP - Multi-Layer Perceptron. AutoAtlas' prediction performance improves with increasing $C_a$ until $C_a=16$.
Performance is measured using the coefficient of determination, $\left(R^2\right)$, (larger is better) and the mean
absolute error (smaller is better). }
\begin{tabular}{|c|c|c|c|c|c|c|c|c|c|c|c|c|c|c|c|c|} 
 \hline
 & \multicolumn{8}{c|}{Prediction of meta-data \textit{Strength-Unadj}} & \multicolumn{8}{c|}{Prediction of meta-data \textit{Strength-AgeAdj}} \\\hline
&  \multicolumn{4}{c|}{Coefficient of determination $\left(R^2\right)$} &   \multicolumn{4}{c|}{Mean Absolute Error} &  \multicolumn{4}{c|}{Coefficient of determination $\left(R^2\right)$} &   \multicolumn{4}{c|}{Mean Absolute Error}\\\hline 
& Lin & NNbor & SVM & MLP & Lin & NNbor & SVM & MLP & Lin & NNbor & SVM & MLP & Lin & NNbor & SVM & MLP \\\hline
FS & 0.343 & 0.323 & 0.387 & 0.383 & 7.02 & 7.03 & 6.83 & 6.75 & 0.350 & 0.324 & 0.385 & 0.384 & 13.14 & 13.11 & 12.18 & 12.51 \\\hline
$AA4$ & 0.371 & 0.318 & 0.378 & 0.392 & 6.96 & 7.30 & 6.94 & 6.82 & 0.378 & 0.322 & 0.380 & 0.393 & 12.68 & 13.45 & 12.56 & 12.55 \\\hline
$AA8$ & 0.402 & 0.382 & 0.421 & 0.397 & 6.77 & \textbf{6.84} & 6.73 & 6.63 & 0.405 & 0.374 & 0.418 & 0.408 & 12.44 & \textbf{12.67} & 12.16 & 12.17 \\\hline
$AA16$ & \textbf{0.451} & \textbf{0.384} & \textbf{0.449} & 0.451 & \textbf{6.53} & 6.87 & 6.58 & 6.51 & \textbf{0.456} & \textbf{0.379} & 0.420 & 0.423 & 12.13 & 12.86 & \textbf{12.07} & 12.13 \\\hline
$AA32$ & 0.435 & 0.335 & 0.439 & \textbf{0.454} & 6.64 & 7.18 & \textbf{6.55} & \textbf{6.47} & 0.442 & 0.335 & 0.398 & \textbf{0.457} & 12.25 & 13.32 & 12.26 & \textbf{11.71} \\\hline
$AA64$ & 0.440 & 0.370 & 0.430 & 0.446 & 6.58 & 7.10 & 6.61 & 6.49 & 0.448 & 0.375 & \textbf{0.421} & 0.453 & \textbf{12.01} & 12.87 & 12.09 & 11.92 \\\hline
\end{tabular}
\end{table*}

\begin{table*}[!htb]
\begin{center}
\caption{\label{tab:labscores}
Importance score for each $16$ dimensional embedding feature $e_i$ (section \ref{sec:replearn}) 
obtained using AutoAtlas with $C_a=16$. 
For each partition $i$, the importance score is computed as the percentage increase 
in the mean absolute error after permutation of the 
embedding features $e_i$ among all the subjects.
From the scores, we can see that a few partitions are consistently determined to be important 
by multiple regressors.
}
\begin{tabular}{c}
(a) Prediction of meta-data \textit{Strength-Unadj}\\
\begin{tabular}{|c|c|c|c|c|c|c|c|c|c|c|c|c|c|c|c|c|} 
 \hline
Optim & $f_{0}$ & $f_{1}$ & $f_{2}$ & $f_{3}$ & $f_{4}$ & $f_{5}$ & $f_{6}$ & $f_{7}$ & $f_{8}$ & $f_{9}$ & $f_{10}$ & $f_{11}$ & $f_{12}$ & $f_{13}$ & $f_{14}$ & $f_{15}$ \\\hline
Lin & 1.33 & 0.74 & 3.57 & 1.24 & 2.15 & 1.39 & 3.09 & 2.58 & 0.81 & 2.41 & 6.20 & -0.01 & 2.06 & 3.57 & 6.18 & -0.34 \\\hline
SVM & 0.75 & 0.11 & 3.41 & 1.13 & 0.91 & -1.27 & 1.01 & 3.46 & 0.28 & -1.05 & 1.61 & -0.17 & -0.34 & 2.69 & 3.28 & 0.63 \\\hline
NNbor & 0.94 & 1.43 & 1.52 & 0.11 & 1.07 & 0.96 & 0.75 & -0.10 & 1.22 & -0.22 & 0.99 & 0.17 & 0.49 & 1.27 & 1.02 & 1.35 \\\hline
MLP & 4.04 & 1.66 & 6.58 & 2.00 & 2.18 & 0.33 & 0.59 & 3.88 & 3.12 & 1.89 & 4.46 & 0.05 & 2.33 & 0.97 & 5.44 & 0.35 \\\hline
\end{tabular}
\vspace{0.05in}\\
(b) Prediction of meta-data \textit{Strength-AgeAdj}\\
\begin{tabular}{|c|c|c|c|c|c|c|c|c|c|c|c|c|c|c|c|c|} 
 \hline
Optim & $f_{0}$ & $f_{1}$ & $f_{2}$ & $f_{3}$ & $f_{4}$ & $f_{5}$ & $f_{6}$ & $f_{7}$ & $f_{8}$ & $f_{9}$ & $f_{10}$ & $f_{11}$ & $f_{12}$ & $f_{13}$ & $f_{14}$ & $f_{15}$ \\\hline
Lin & 1.18 & 0.88 & 4.37 & 1.54 & 2.21 & 2.40 & 3.39 & 3.98 & 1.08 & 3.33 & 5.14 & 0.11 & 2.33 & 3.38 & 5.61 & -0.15 \\\hline
SVM & 1.60 & 0.81 & 6.43 & 1.57 & 1.18 & -0.55 & 1.11 & 3.67 & 0.88 & -0.55 & 2.30 & 0.08 & -0.01 & 1.69 & 3.08 & 0.06 \\\hline
NNbor & 0.87 & 1.06 & 1.54 & -0.02 & 0.87 & 0.47 & 0.65 & 0.07 & 0.79 & -0.29 & 0.96 & -0.11 & 0.35 & 0.79 & 0.69 & 0.74 \\\hline
MLP & 2.08 & 0.56 & 7.58 & 0.47 & 1.70 & 1.14 & 0.42 & 4.74 & 0.15 & 0.67 & 2.43 & 0.07 & 0.82 & 0.46 & 3.05 & -0.06 \\\hline
\end{tabular}
\end{tabular}
\end{center}
\end{table*}

\begin{figure}[!htb]
\begin{center}
Prediction of meta-data \textit{Strength-Unadj}
\begin{tabular}{ccc}
\hspace{-0.22in}
\includegraphics[width=1.2in]{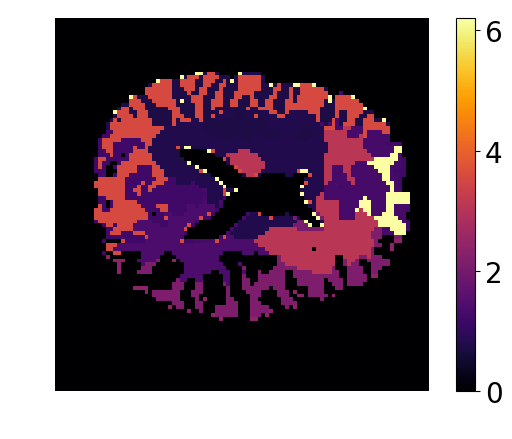} &
\hspace{-0.22in}
\includegraphics[width=1.2in]{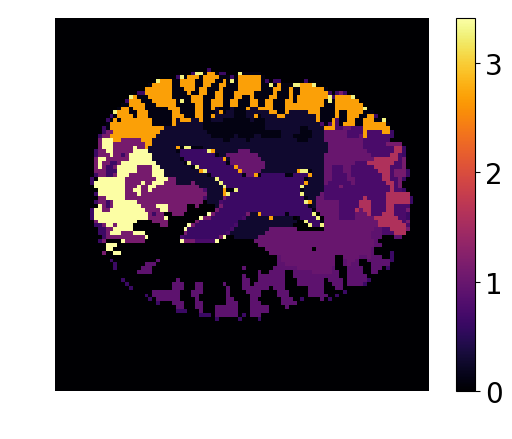} &
\hspace{-0.22in}
\includegraphics[width=1.2in]{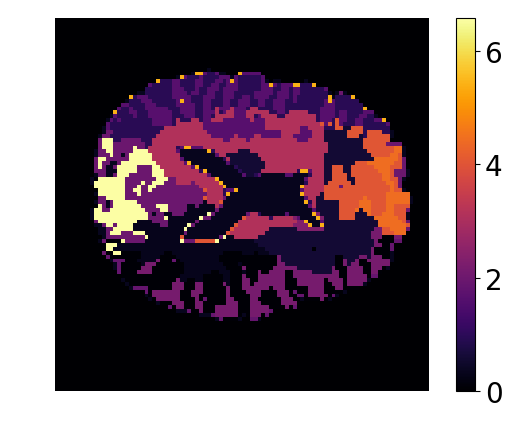} 
\vspace{-0.12in}\\
\hspace{-0.22in} (a) Lin & \hspace{-0.22in} (b) SVM & \hspace{-0.22in} (c) MLP\\
\hspace{-0.22in}
\includegraphics[width=1.2in]{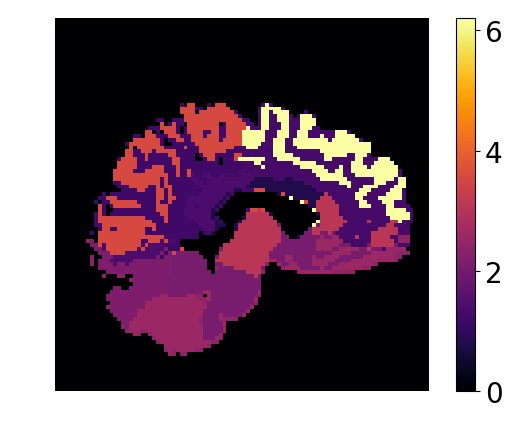} &
\hspace{-0.22in}
\includegraphics[width=1.2in]{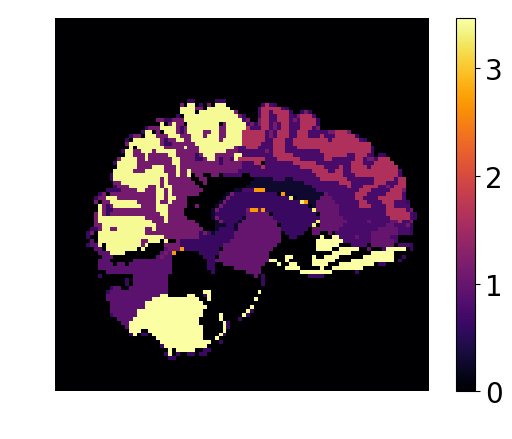} &
\hspace{-0.22in}
\includegraphics[width=1.2in]{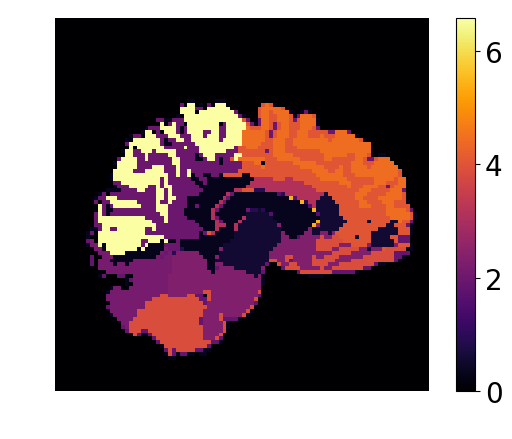} 
\vspace{-0.12in}\\
\hspace{-0.22in} (d) Lin & \hspace{-0.22in} (e) SVM & \hspace{-0.22in} (f) MLP\\
\hspace{-0.22in}
\includegraphics[width=1.2in]{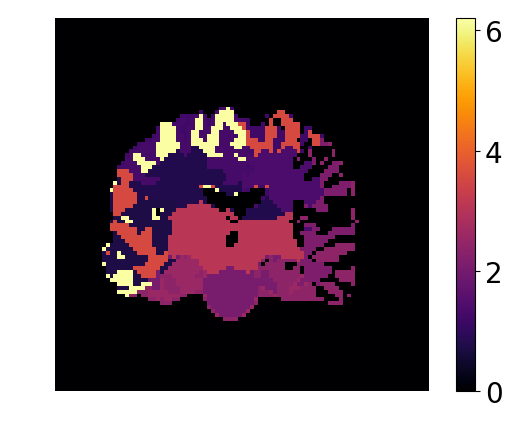} &
\hspace{-0.22in}
\includegraphics[width=1.2in]{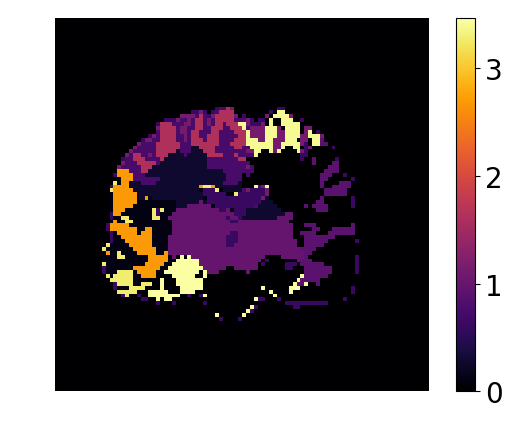} &
\hspace{-0.22in}
\includegraphics[width=1.2in]{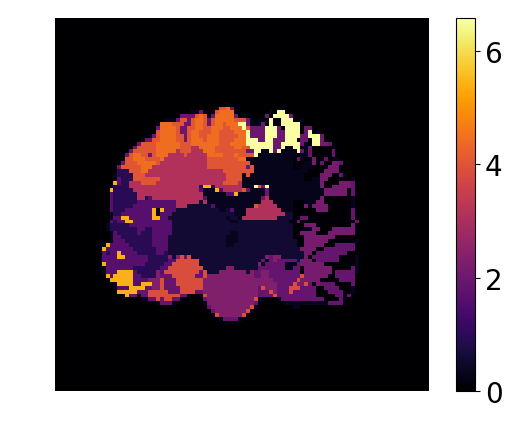} 
\vspace{-0.12in}\\
\hspace{-0.22in} (g) Lin & \hspace{-0.22in} (h) SVM & \hspace{-0.22in} (i) MLP\\
\end{tabular}\\
\vspace{0.05in}
Prediction of meta-data \textit{Strength-AgeAdj}\\
\begin{tabular}{ccc}
\hspace{-0.22in}
\includegraphics[width=1.2in]{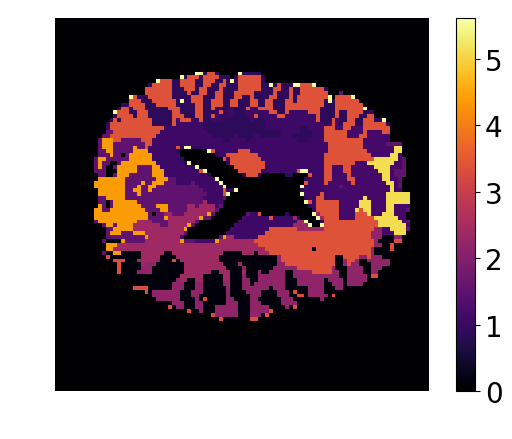} &
\hspace{-0.22in}
\includegraphics[width=1.2in]{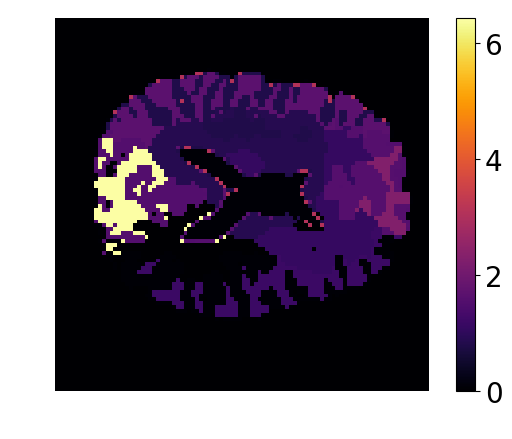} &
\hspace{-0.22in}
\includegraphics[width=1.2in]{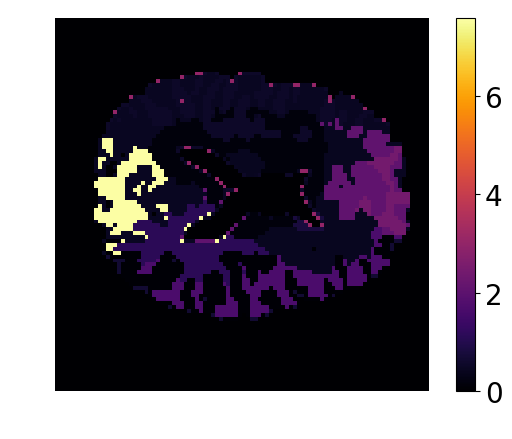} 
\vspace{-0.12in}\\
\hspace{-0.22in} (j) Lin & \hspace{-0.22in} (k) SVM & \hspace{-0.22in} (l) MLP\\
\hspace{-0.22in}
\includegraphics[width=1.2in]{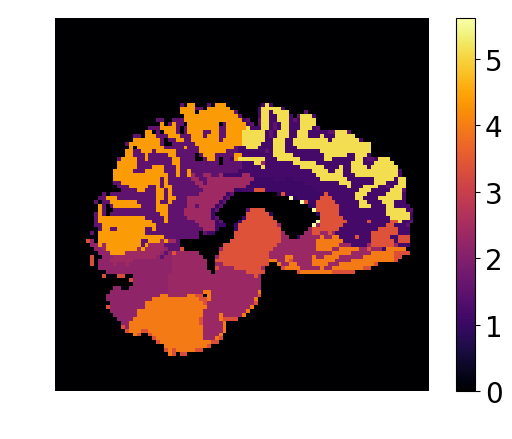} &
\hspace{-0.22in}
\includegraphics[width=1.2in]{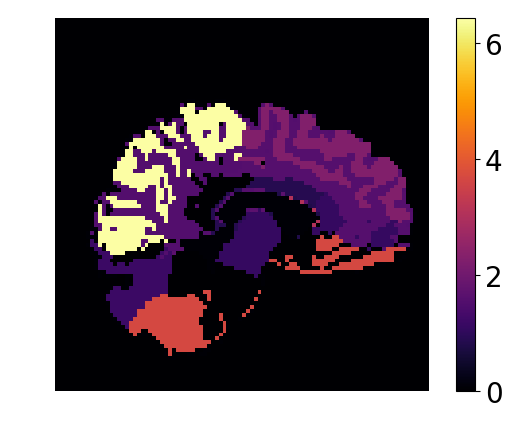} &
\hspace{-0.22in}
\includegraphics[width=1.2in]{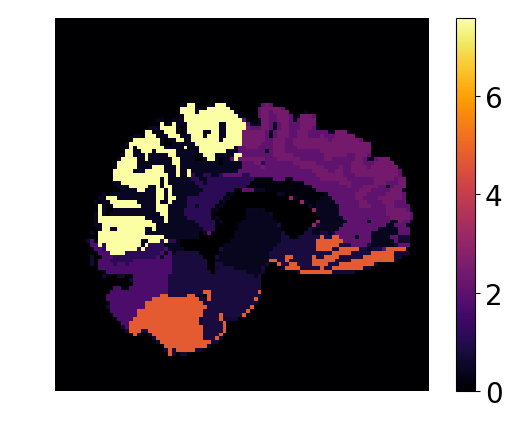} 
\vspace{-0.12in}\\
\hspace{-0.22in} (m) Lin & \hspace{-0.22in} (n) SVM & \hspace{-0.22in} (o) MLP\\
\hspace{-0.22in}
\includegraphics[width=1.2in]{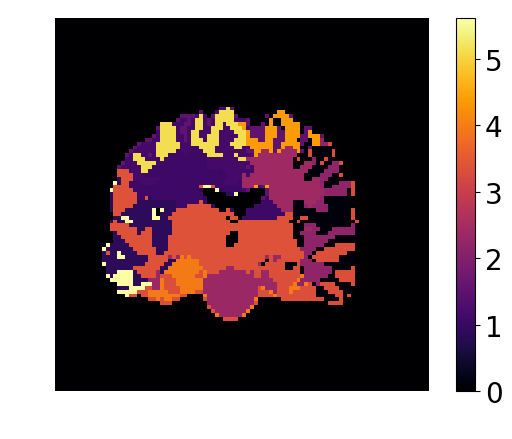} &
\hspace{-0.22in}
\includegraphics[width=1.2in]{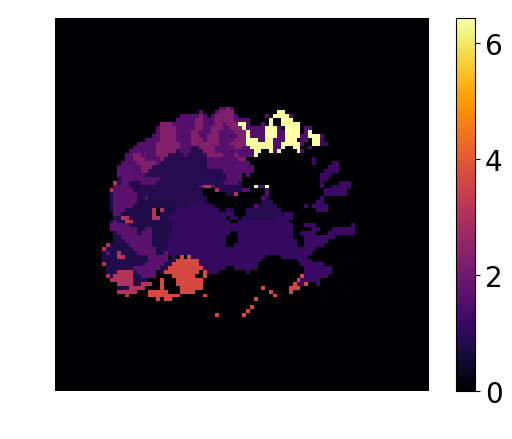} &
\hspace{-0.22in}
\includegraphics[width=1.2in]{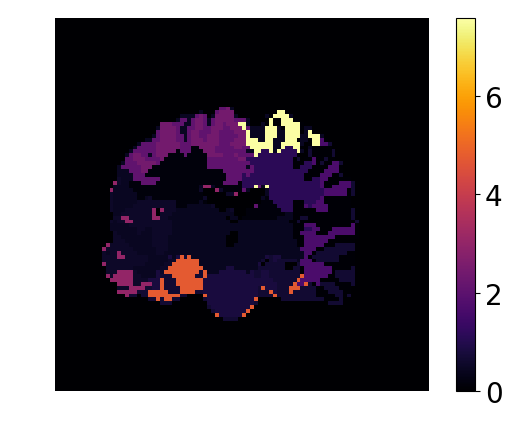} 
\vspace{-0.12in}\\
\hspace{-0.22in} (p) Lin & \hspace{-0.22in} (q) SVM & \hspace{-0.22in} (r) MLP\\
\end{tabular}
\end{center}
\caption{\label{fig:aarank} Overlay map of the importance scores on the brain volume of an individual subject. Partitions located in the cortical surface seem to be more important than the interior partitions when predicting both \textit{Strength-Unadj} and \textit{Strength-AgeAdj}.
}
\end{figure}

\begin{table*}[h!]
\caption{\label{tab:ablrl} Ablation study of AutoAtlas losses by comparing the prediction performance during regression of \textit{Strength-Unadj}.
The best performance is obtained along the last row where all the three losses are in use ($\lambda_{AD}\neq 0$, $\lambda_{NLS}\neq 0$, and $\lambda_{RE}\neq 0$).
In the absence of AD loss ($\lambda_{AD}=0$), we see that AutoAtlas only uses one label to represent the whole brain resulting in increased root mean squared error (RMSE) for the reconstruction.
In the absence of NLS loss ($\lambda_{NLS}=0$), even though the RMSE is the lowest, the non-smooth partitions (Fig. \ref{fig:ablsegc}(a)) result in lower prediction performance than the last row. 
Without the RE loss ($\lambda_{RE}=0$), the prediction performance suffers due to loss of information in the embeddings $e_i$.}
\begin{center}
\begin{tabular}{|c|c|c|c|c|c|c|c|c|c|c|c|c|} 
\hline
\multicolumn{5}{|c|}{} &  \multicolumn{4}{c|}{Coefficient of determination $\left(R^2\right)$} &   \multicolumn{4}{c|}{Mean absolute error} \\\hline 
$\lambda_{RE}$ & $\lambda_{NLS}$ & $\lambda_{AD}$ & RMSE & Regions & Lin & SVM & NNbor & MLP & Lin & SVM & NNbor & MLP\\\hline
0.0 & 0.005 & 0.1 & 0.60 & 12 & 0.37 & 0.35 & 0.35 & -0.01 & 7.12 & 7.18 & 7.26 & 9.23\\\hline
1.0 & 0.0 & 0.1 & 0.02 & 16 & 0.38 & 0.41 & 0.39 & 0.33 & 7.03 & 6.76 & 6.93 & 7.20\\\hline
1.0 & 0.005 & 0.0 & 0.09 & 1 & 0.38 & 0.38 & 0.35 & -0.00 & 6.90 & 6.92 & 7.00 & 9.21\\\hline
1.0 & 0.005 & 0.1 & 0.04 & 16 & 0.45 & 0.45 & 0.38 & 0.45 & 6.53 & 6.58 & 6.87 & 6.51\\\hline
\end{tabular}
\end{center}
\end{table*}

\begin{figure*}[!htb]
\begin{center}
\begin{tabular}{ccccc}
{\small Partitions satisfying the} & {\small Partitions with $5.625\%$} & {\small RMSE between} & {\small Percentage of voxels} & {\small Partition smoothness}\\
{\small AD constraint in Eq. \eqref{eq:adconspart}} & {\small or more brain voxels} & {\small $\hat{x}$ and $x$} & {\small that satisfy Eq. \eqref{eq:confvoxs}} & {\small given by Eq. \eqref{eq:smoothmeas}} \\
\hspace{-0.22in}
\includegraphics[width=1.4in]{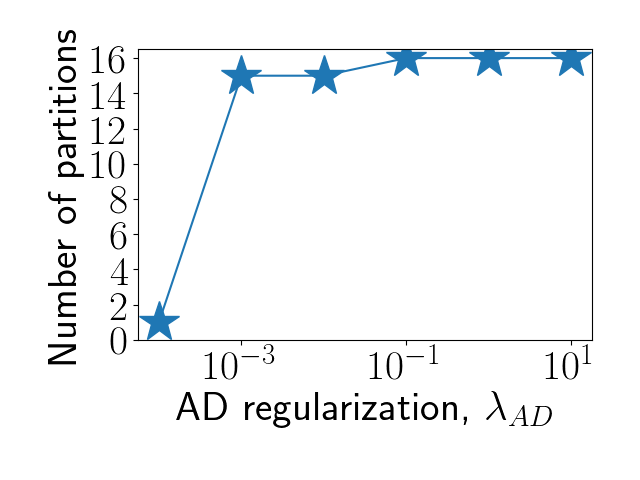} &
\hspace{-0.22in}
\includegraphics[width=1.4in]{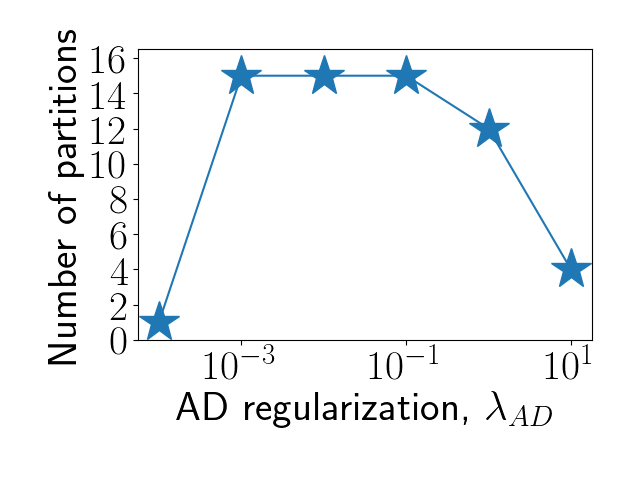} &
\hspace{-0.22in}
\includegraphics[width=1.4in]{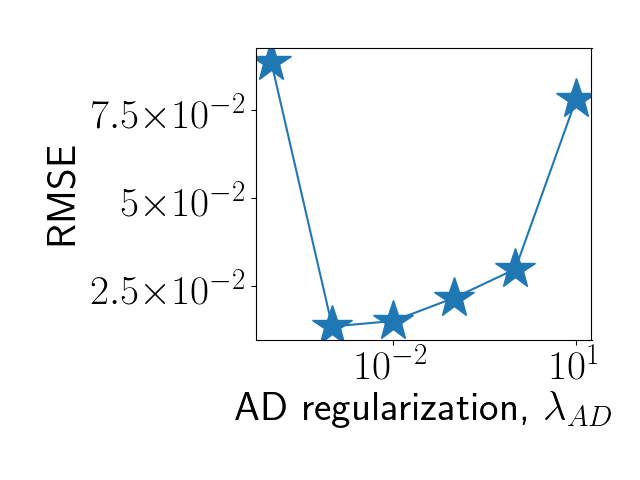} &
\hspace{-0.22in}
\includegraphics[width=1.4in]{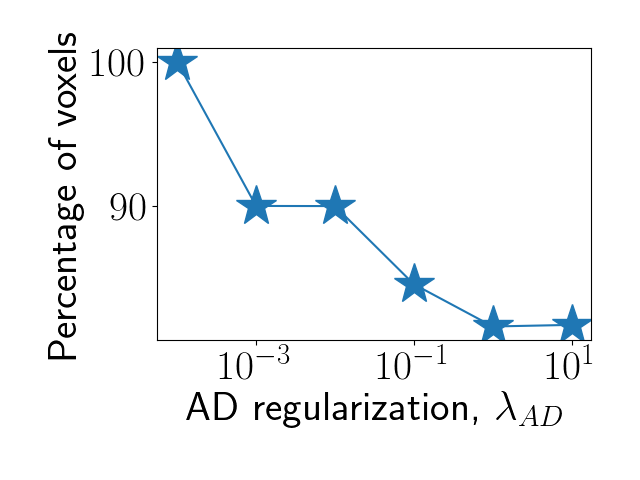} &
\hspace{-0.22in}
\includegraphics[width=1.4in]{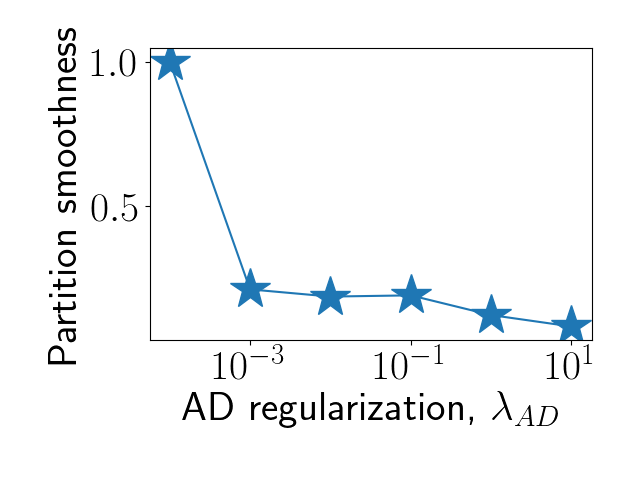} \vspace{-0.12in}\\
(a) &  (b) &  (c) &  (d) & (e) \\
\includegraphics[width=1.4in]{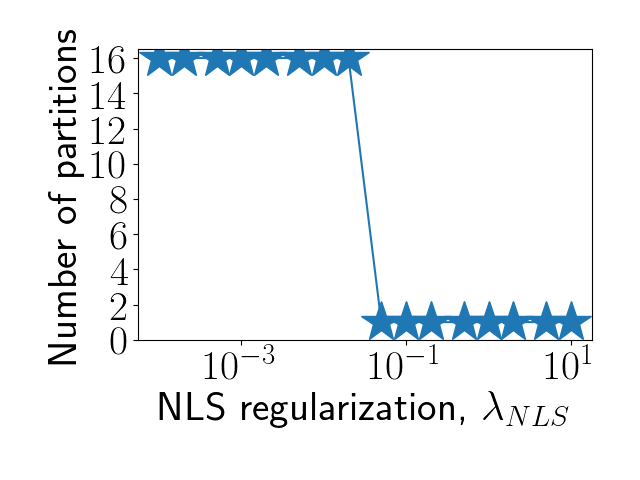} &
\hspace{-0.22in}
\includegraphics[width=1.4in]{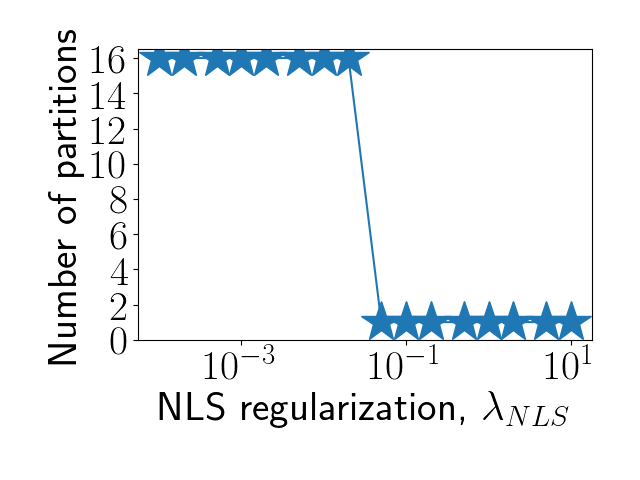} &
\hspace{-0.22in}
\includegraphics[width=1.4in]{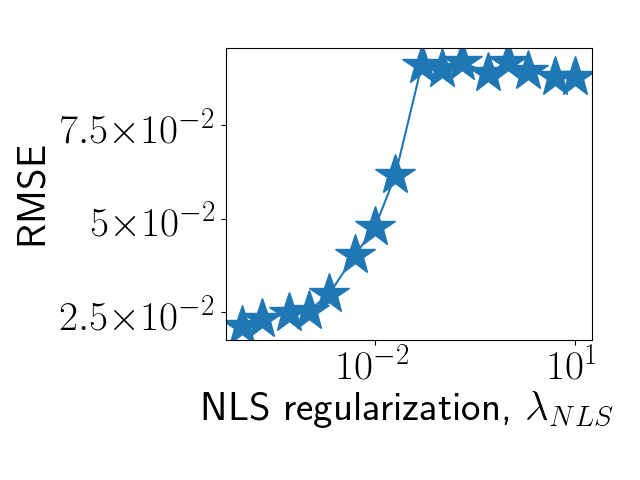} &
\hspace{-0.22in}
\includegraphics[width=1.4in]{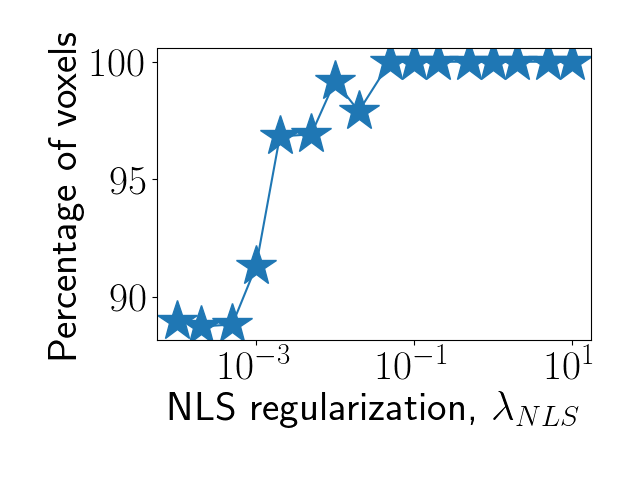} &
\hspace{-0.22in}
\includegraphics[width=1.4in]{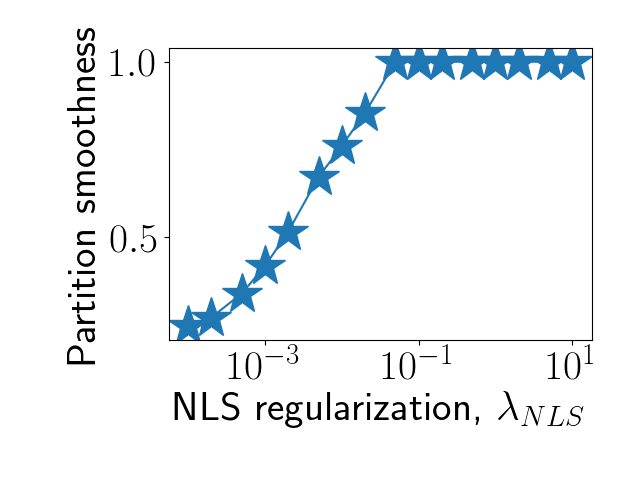} \vspace{-0.12in}\\
(f) &  (g) & (h) & (i) & (j) \\
\end{tabular}
\end{center}
\caption{\label{fig:swpablstudy} Partitioning performance and reconstruction fidelity as a function of the regularization hyper-parameters in equation \eqref{eq:totloss}. 
In the $1^{st}$ column, we plot the number of partitions that satisfy the AD constraint in equation \eqref{eq:adconspart}. In the $2^{nd}$ column, we plot the number of partitions that occupy more than $5.625\%$ of all voxels inside the brain, where the partition associated with each voxel is given by equation \eqref{eq:partpred}. In the $3^{rd}$ column, we plot the RMSE between the reconstruction, $\hat{x}$, and the ground-truth volume $x$ (Fig. \ref{fig:fullblkdiag}). In the $4^{th}$ column, we plot the percentage of voxels that predict only one partition label with high confidence (equation \eqref{eq:confvoxs}). In the $5^{th}$ column, we plot a measure of the smoothness of the AutoAtlas partitions as quantified by equation \eqref{eq:smoothmeas}. In the first row, we plot as a function of the AD regularization hyper-parameter $\lambda_{AD}$ when $\lambda_{NLS}=0$. In the second row, we plot as a function of the NLS regularization hyper-parameter $\lambda_{NLS}$ when $\lambda_{AD}=0.1$. Note that Eq. is an abbreviation for equation.}
\end{figure*}

\begin{figure}[!htb]
\begin{center}
\begin{tabular}{ccc}
\hspace{-0.22in}
\includegraphics[width=1.2in]{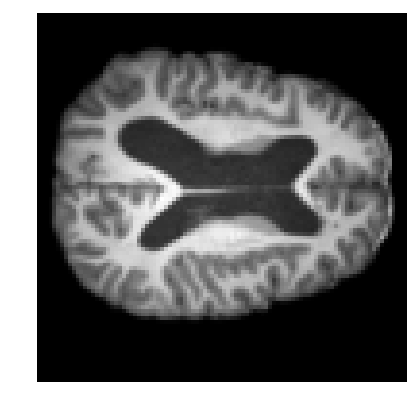} &
\hspace{-0.22in}
\includegraphics[width=1.2in]{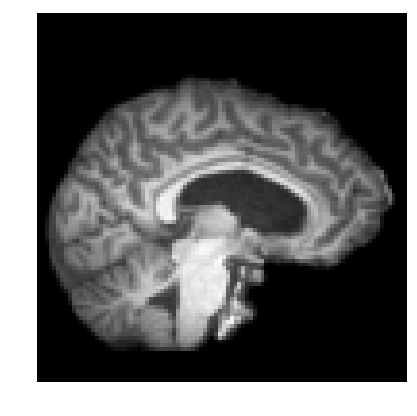} &
\hspace{-0.22in}
\includegraphics[width=1.2in]{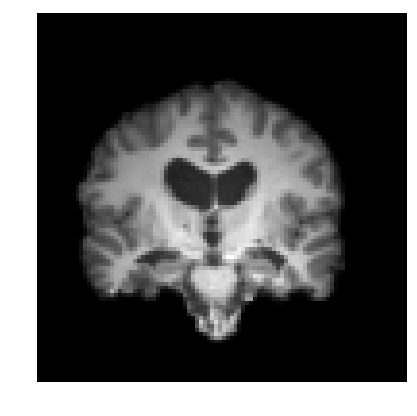} 
\vspace{-0.12in}\\
\hspace{-0.22in} (a)  & \hspace{-0.22in} (b) & \hspace{-0.22in} (c) \\
\hspace{-0.22in}
\includegraphics[width=1.2in]{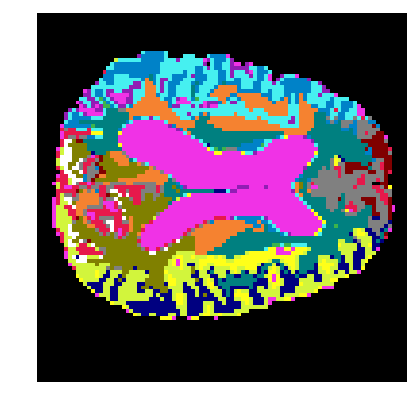} &
\hspace{-0.22in}
\includegraphics[width=1.2in]{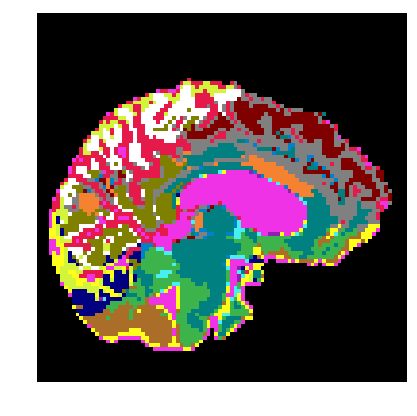} &
\hspace{-0.22in}
\includegraphics[width=1.2in]{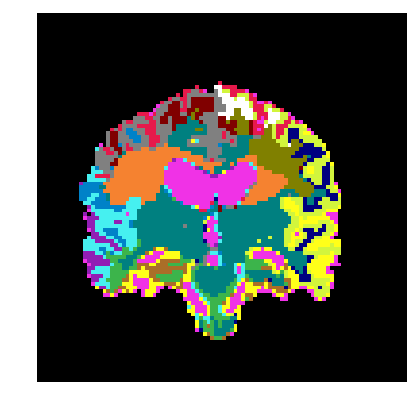} 
\vspace{-0.12in}\\
\hspace{-0.22in} (d)  & \hspace{-0.22in} (e) & \hspace{-0.22in} (f) \\
\end{tabular}
\end{center}
\caption{\label{fig:costtbivol} 
Testing of AutoAtlas on an MRI of a subject with TBI. (a,b,c) show 
perpendicularly oriented slices through the center of the MRI scan. (d,e,f) show the
partitions generated using AutoAtlas.
In the slice shown in (d), the ventricles appear enlarged when compared to 
the MRIs of healthy subjects in Fig. \ref{fig:gtseg_subj}.
}
\end{figure}

Table \ref{tab:predscores} compares the predictive performance of AutoAtlas' embedding features
and Freesurfer features as measured using the coefficient of 
determination\footnote{The $R^2$ score is one for a regressor that always predicts correctly.} 
($R^2$) and mean absolute 
error\footnote{Mean of the absolute value differences between prediction and ground-truth} metrics.
We used the python based machine learning package \textit{scikit-learn} \cite{scikit-learn} to train and test all the regressors in Tables \ref{tab:predscores}, \ref{tab:labscores}, and \ref{tab:ablrl}. 
The regularization for each predictor was tuned using an automated grid 
search strategy and 3-fold cross validation on the training data.
More details regarding the chosen hyper-parameters 
for the predictors in Table \ref{tab:predscores}
are presented in Appendix \ref{app:rlpars}.
The prediction performance using AutoAtlas features
generally improve with increasing values for the embedding size $C_a$ 
as shown in Table \ref{tab:predscores}.
This behavior can be attributed to the improved information retention when using larger embedding sizes.
However, the performance stagnates beyond $C_a=16$ since increasing $C_a$ beyond
$16$ does not further improve the encoded information content in the embedding.

In Table \ref{tab:labscores}, we quantify the importance 
of each individual embedding feature vector $e_i$ (section \ref{sec:replearn}),
which encodes the information within partition $i$,
when predicting \textit{Strength-Unadj} and \textit{Strength-AgeAdj}.
The importance score for a partition $i$ is computed as the percentage increase in 
mean absolute error after randomly permuting\footnote{Average score is reported over $10$ randomized permutations.} the embedding features $e_i$.
Permutation refers to the process of randomly shuffling the vector of embedding features 
among all the subjects in the test set.
In Fig. \ref{fig:aarank}, we overlay the importance score for $e_i$ (Table \ref{tab:labscores})
over partition $i$
in order to associate the partition-specific importance scores 
with the corresponding location of each partition in the MRI scan.
From Fig. \ref{fig:aarank}, we can see that 
partitions located along the cortical surface seem to be more important than the interior regions
when predicting \textit{Strength-Unadj} and \textit{Strength-AgeAdj}. 
We also include supporting multimedia files in the form
of MP4 videos to visualize the MRI volumes, AutoAtlas partitions, 
and the overlay of importance scores on the partitions.

Table \ref{tab:ablrl} is an ablation study of the
RE (equation \eqref{eq:relloss}), NLS (equation \eqref{eq:nsslloss}), 
and AD (equation \eqref{eq:adlloss}) losses
that compares the representation learning performance. 
The $5^{th}$ column (titled ``Regions") in Table \ref{tab:ablrl}
shows that both AD loss and RE loss play a pivotal role in ensuring that all $16$ labels are 
utilized to represent the volume. 
From $4^{th}$ column (titled ``RMSE") of Table \ref{tab:ablrl},
we see that while the constraints imposed by the NLS loss tend to increase the reconstruction error,
AD and RE losses strive to reduce it.  
Irrespective of the reconstruction error, the simultaneous use of 
the RE, NLS, and AD losses results in the best representation learning performance.

To guide our selection of regularization hyper-parameters, 
$\lambda_{NLS}$ and $\lambda_{AD}$ in equation \eqref{eq:totloss}, 
we plot various metrics that quantify the nature of AutoAtlas partitions and reconstruction 
on the training set as a function of either $\lambda_{NLS}$ or $\lambda_{AD}$ in Fig. \ref{fig:swpablstudy}.
Due to underfitting of our AutoAtlas models, 
we chose to perform hyper-parameter tuning using the training set 
instead of a separate validation set.
For the purpose of hyper-parameter tuning, we only trained AutoAtlas for 200 epochs.
Each plot in Fig. \ref{fig:swpablstudy} perform a parameter sweep along the horizontal axis 
over either $\lambda_{AD}$ or $\lambda_{NLS}$ in logarithmic steps 
while assuming constant values for the other regularization hyper-parameters.
Fig. \ref{fig:swpablstudy} (a, f) 
plot the number of partitions that satisfy the AD constraint 
shown in equation \eqref{eq:adconspart} along the vertical axis.
Fig. \ref{fig:swpablstudy} (b, g) plot the number of partitions whose voxel count is more than 
$5.625\%$\footnote{Given by $u_i\times 100=5.625\%$, where $u_i$ is defined in 
equation \eqref{eq:adconspart}.} of the total number of voxels within the brain 
(a voxel's partition membership is defined in equation \eqref{eq:partpred}).  
Fig. \ref{fig:swpablstudy} (c, h) plot the root mean squared error (RMSE) 
between the reconstruction, $\hat{x}$ as defined in equation \eqref{eq:mrirec}, 
and the ground-truth input volume $x$ (Fig. \ref{fig:fullblkdiag}). 
Fig. \ref{fig:swpablstudy} (d, i) show the percentage of voxels that satisfy
the constraint $\max_i y_{i,j} \geq 0.99$, which indicates high-confidence predictions 
(i.e., satisfy equation \eqref{eq:confvoxs}).
Lastly, Fig. \ref{fig:swpablstudy} (e, j) plot a measure of 
the smoothness of AutoAtlas partitions as defined in equation \eqref{eq:smoothmeas}. 
Note that all plots in Fig. \ref{fig:swpablstudy} 
plot the average value over all volumes in the training set along the vertical axis. 

For hyper-parameter tuning, we first empirically choose an appropriate  
$\lambda_{AD}$ while setting $\lambda_{RE}=1$ and $\lambda_{NLS}=0$ in equation \eqref{eq:totloss}.
In Fig. \ref{fig:swpablstudy} (a-e), we plot metrics that quantify the partitioning performance 
and reconstruction fidelity as a function of $\lambda_{AD}$ when $\lambda_{NLS}=0.0$.
From Fig. \ref{fig:swpablstudy} (a), we can see that $\lambda_{AD}$ must be chosen to 
have a sufficiently large value in order to ensure that the AD constraint is satisfied.
However, the number of partitions with significant voxel count 
shown in Fig. \ref{fig:swpablstudy} (b) does not follow a predictable pattern due to the absence of NLS loss.
From Fig. \ref{fig:swpablstudy} (c), we observe that the RMSE measure 
appears to be inversely correlated to the number of partitions in Fig. \ref{fig:swpablstudy} (b).
As $\lambda_{AD}$ is increased, we also observe a decrease in both the 
percentage of voxels with high-confidence predictions in Fig. \ref{fig:swpablstudy} (d)
and the measure of partition smoothness in  Fig. \ref{fig:swpablstudy} (e).
Guided by Fig. \ref{fig:swpablstudy} (a-e), we choose $\lambda_{AD}=0.1$ 
since it is the lowest value that satisfies the AD constraint in Fig. \ref{fig:swpablstudy} (a).

To choose an appropriate value for $\lambda_{NLS}$, 
we examine Fig. \ref{fig:swpablstudy} (f-j) that analyze the partitioning performance
and reconstruction fidelity as a function of $\lambda_{NLS}$ when $\lambda_{RE}=1$ and $\lambda_{AD}=0.1$.
From Fig. \ref{fig:swpablstudy} (f, g), 
we observe that large values for $\lambda_{NLS}$
invalidate the AD constraint by forcing the whole brain to be represented by only one partition.
Unlike Fig. \ref{fig:swpablstudy} (a, b), Fig. \ref{fig:swpablstudy} (f) and Fig. \ref{fig:swpablstudy} (g)
follow a similar behavior due to the presence of NLS loss.
For large values of $\lambda_{NLS}$ that result in a single partition 
for the whole brain (Fig. \ref{fig:swpablstudy} (f, g)),
the RMSE in Fig. \ref{fig:swpablstudy} (h) is also high since a single
low-capacity autoencoder is tasked with reconstruction of the whole brain.
As $\lambda_{NLS}$ is increased, we also observe an increase in the
percentage of voxels with high-confidence predictions in Fig. \ref{fig:swpablstudy} (i)
and the measure of smoothness in Fig. \ref{fig:swpablstudy} (j).
We choose a value of $\lambda_{NLS}=0.005$ that provides an acceptable trade-off
of low RMSE, smooth partitions, and a large number of voxels with high-confidence predictions. 
While we chose to avoid further tuning of $\lambda_{AD}$, 
such a tuning may be valuable in a different application. 

To evaluate the robustness of AutoAtlas and its applicability to a broad range of datasets, 
we tested our AutoAtlas model (the model used in Fig. \ref{fig:gtseg_subj} and Fig. \ref{fig:gtseg_zyx}) 
on the MRI of a subject diagnosed with Traumatic Brain Injury (TBI) \cite{costtbi}. 
Note that AutoAtlas was trained solely on MRIs of healthy subjects 
in the HCP-YA dataset \cite{Van_Essen2013-xf, Glasser2016-kr, Smith2013-vw, Barch2013-wy, Van_Essen2012-cf, glasser2013minimal}. 
Hence, testing on an MRI with TBI represents a drastic change
from the distribution of samples used to train AutoAtlas.
A comparison between various orthogonal slices through the MRI volume
and its AutoAtlas partitions is shown in Fig. \ref{fig:costtbivol}.
In spite of TBI, the performance of 
AutoAtlas in Fig. \ref{fig:costtbivol} is consistent with Fig. \ref{fig:gtseg_subj}, 
Fig. \ref{fig:gtseg_zyx}, and Fig. \ref{fig:colleg}. 
This behavior demonstrates the robust nature of AutoAtlas
even when evaluated on a radically different dataset than the one used for training.

Our approach is also applicable to the analysis 
of volumes containing pathological features. 
Depending on the underlying nature of the pathology, 
we may observe different characteristics for the partitions 
and the embeddings. 
For subjects with traumatic brain injury, 
we may see partitions that are either entirely missing 
or significantly different from normative cases. 
Alternatively, Alzheimer's Disease may result in 
volumetric changes to the partitions due to
 reduced gray matter volume. 
Since the embedding features (equation \eqref{eq:ftall}) 
and the partitions' shapes and sizes are directly 
impacted by the changes described above, 
we expect that the AutoAtlas representation 
may be useful in prediction and localization of pathology.

\section{Acknowledgment}
LLNL-JRNL-815848 and LLNL-VIDEO-816094. This work was performed under the auspices of the 
U.S. Department of Energy by Lawrence Livermore National 
Laboratory (LLNL) under Contract DE-AC52-07NA27344.
This work has been supported in part by grants from the DOE Office of Science’s Advanced Computing Science Research (Grant \#KJ0403020).
Data was provided by the Human Connectome Project (HCP), WU-Minn Consortium (Principal Investigators: David Van Essen and Kamil Ugurbil; 1U54MH091657) funded by the 16 NIH Institutes and Centers that support the NIH Blueprint for Neuroscience Research; and by the McDonnell Center for Systems Neuroscience at Washington University.

\appendices
\section{Neural Network Architecture}
\label{app:nnarch}

\begin{figure}[!tb]
    \begin{center}
    \hspace{-0.2in}
    \includegraphics[width=3.6in,keepaspectratio=true]{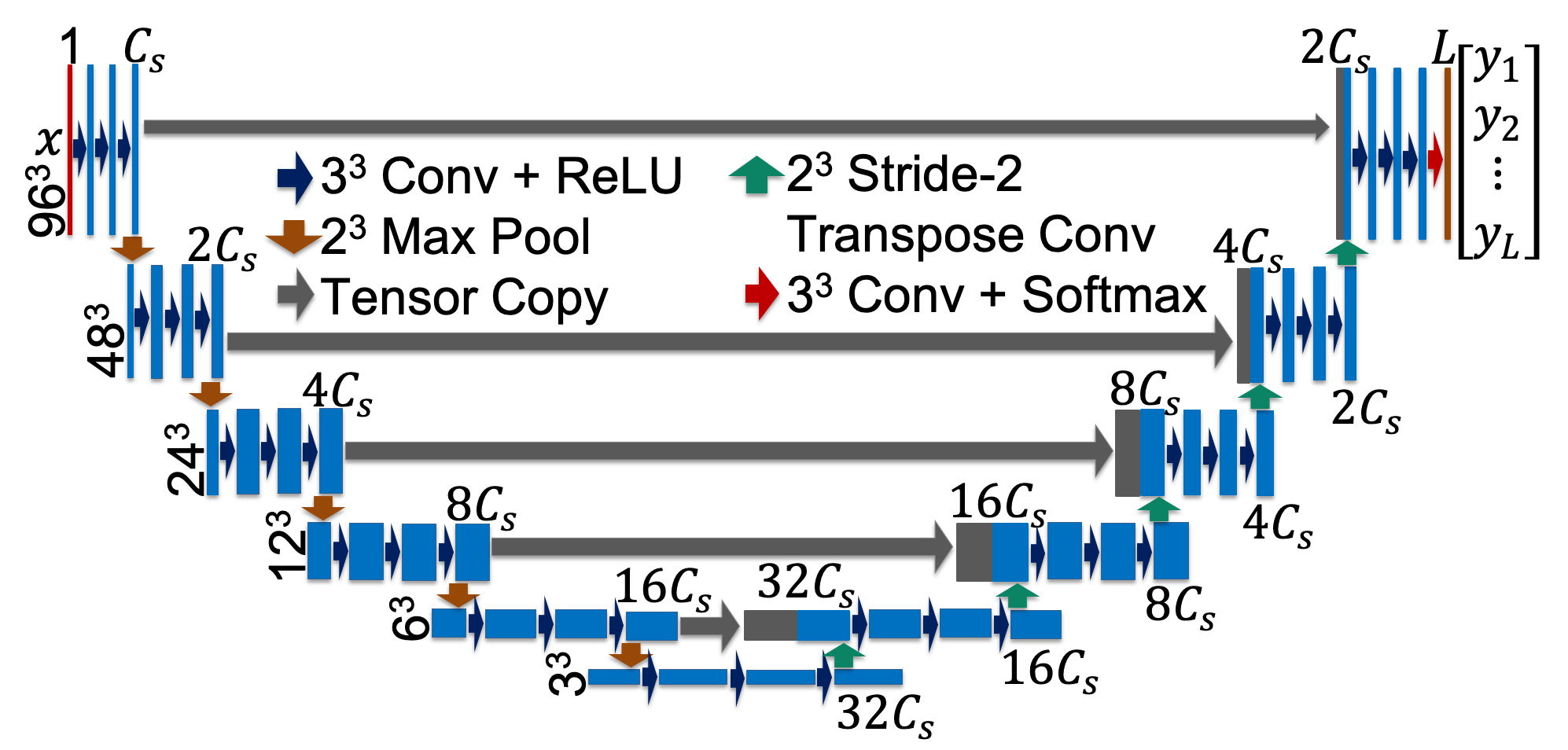}
    \end{center}
    \caption{\label{fig:UNet}
	A 3D U-Net is used to partition the input volume into $L$ partitions.
	Each rectangle represents a tensor. The number of channels is indicated above or below the tensor. 
 The spatial size is indicated to the left of the tensor.
    }
\end{figure}

The architecture of U-net used in this paper is shown in Fig. \ref{fig:UNet}.
It consists of a contracting path where the 3D size is progressively halved in each dimension
 followed by an expansive path where the 3D size is progressively doubled. 
The contracting path consists of multiple downsampling stages where each 
stage has three 3D convolutional layers each of which is followed by a ReLU activation layer. 
The first convolutional layer at each stage doubles the number of channels when compared to its input.
The output of every stage is downsampled by a factor of $2$ in each spatial dimension 
using a max-pool layer.
The number of channels at the output of the first convolutional layer is denoted by the parameter $C_s$. 
At the end of the contracting path, there is an expansive path with multiple stages 
each of which consists of three sets of convolutional and ReLU layers.
The input for each stage in the expansive path is the concatenation of the output of the previous stage and 
the same sized output from the earlier contracting path. 
The output of each stage is up-sampled by a factor of $2$ along each spatial dimension 
using a 3D transposed convolutional layer with a stride of $2$.
This layer also halves the number of channels when compared to its input.
The last layer of the U-net is a softmax over $L$ number of channels 
for each voxel. 
The softmax computes the classification probability vector for each voxel. 

Similar to the U-net, our autoencoder architecture also has a contracting path followed by an expansive path.
However, unlike U-net, there are no cross-connections from the contracting path to the expansive path.
This ensures that all information has to flow through the bottleneck layer at the center of the autoencoder 
in between the contracting and expansive paths.
At each stage in the contracting path, the 3D size is halved in each spatial dimension
using a convolutional layer with a stride of $2$. 
In the expansive path, each stage consists of a transposed convolutional layer with a stride of $2$ 
to double the 3D size in each spatial dimension. 
Every convolutional and transposed convolution layers are followed by a ReLU activation layer 
except at the last stage that produces the reconstruction $z_i$.
The number of channels is fixed at a value of $C_a$ at the output of each convolutional layer in the autoencoder.
At the bottleneck layer that is in between the contracting and expansive paths, 
a fully connected layer (FC) transforms the output of the last contracting stage 
into a vector of size $C_a$. 
Another FC layer transforms this vector into a multi-dimensional tensor with $C_a$ number of channels
that serves as input to the first stage in the expansive path.

\section{Representation Learning Parameters}
\label{app:rlpars}

All the machine learning approaches used to predict the meta data $\textit{Strength-Unadj}$ and $\textit{Strength-AgeAdj}$ in Tables \ref{tab:predscores} and \ref{tab:labscores}
were implemented using the python package \textit{scikit-learn}\footnote{https://scikit-learn.org/} version \textit{0.22.1}.
Before training and testing, the class \textit{sklearn.preprocessing.StandardScaler}
was used to normalize all features by mean subtraction and dividing by the standard deviation.
The mean and standard deviation parameters were computed from the training set.

The class \textit{sklearn.linear\_model.Ridge}
was used to train the linear ridge regressor (tag ``Lin" in Tables \ref{tab:predscores} and \ref{tab:labscores}). 
The optimal regularization parameter, $\alpha$, 
was chosen to maximize the $R^2$ score using 3-fold cross-validation 
by searching over a grid of $44$ values, which were logarithmically spaced from a starting value of
$10^{-5}$ up to a maximum value that was strictly less than $10^{6}$.
We used the \textit{lsqr} solver and default values for all 
other parameters.	

The class \textit{sklearn.neighbors.KNeighborsRegressor}
was used to train the nearest neighbor regressor  
(tag ``NNbor" in Tables \ref{tab:predscores} and \ref{tab:labscores}).
The number of neighbors parameter, \textit{n\_neighbors},
was chosen to maximize the $R^2$ score using $3$-fold cross-validation and grid-search
over grid values of $1$, $2$, $3$, $4$, $5$, $6$, $8$, $9$, $11$, 
$13$, $16$, $19$, $22$, $26$, $32$, $38$, $45$, $53$, $64$, $76$, $90$, $107$, $128$, $152$, $181$, and $215$.

The class \textit{sklearn.svm.SVR} was used to 
perform support vector regression (tag ``SVM" in Tables \ref{tab:predscores} and \ref{tab:labscores}).
The regularization parameter $C$ was 
chosen to maximize the $R^2$ score using $3$-fold cross-validation by 
searching over a grid of $44$ values, which were logarithmically spaced from a starting value of
$10^{-5}$ up to a maximum value that was strictly less than $10^{6}$.
The kernel coefficient \textit{gamma} was chosen to be
inversely proportional to the number of features.
All other hyper-parameters were left at their respective default values.

The class \textit{sklearn.neural\_network.MLPRegressor}
was used to perform regression using a multi-layer
perceptron (MLP) architecture (tag ``MLP" in Tables \ref{tab:predscores} and \ref{tab:labscores}).
MLP used two hidden layers of sizes $4$ (first hidden layer) and $2$ (second hidden layer) for all prediction tasks irrespective of the number of features.
We used the \textit{lbfgs} solver and a maximum number of  $10^5$ iterations.
The L2 regularization parameter $\alpha$ was chosen to maximize the $R^2$ score 
using $3$-fold cross-validation by 
searching over a grid of $44$ values, which were logarithmically spaced from a starting value of
$10^{-5}$ up to a maximum value that was strictly less than $10^{6}$.

\bibliographystyle{IEEEtran}
\bibliography{IEEEabrv,paper}


\end{document}